\documentclass[twocolumn,
showpacs,
showkeys,
preprintnumbers,
nofootinbib,
superscriptaddress,
amsmath,
amssymb,
floatfix,
secnumarabic,
aps,
pra,
a4paper,
notitlepage,
final,
]{revtex4-1}%

\usepackage[colorlinks=true,urlcolor=blue]{hyperref}
\usepackage{graphicx}
\usepackage{epsfig}

\newcommand{\beq}{\begin{equation}}
\newcommand{\eeq}{\end{equation}}
\newcommand{\beqar}{\begin{eqnarray}}
\newcommand{\eeqar}{\end{eqnarray}}
\newcommand{\bal}{\begin{aligned}}
\newcommand{\eal}{\end{aligned}}

\def\dalam{\hbox
{\vrule\vbox{\hrule\hbox to 1ex{ \hfill}\kern 1 ex\hrule}\vrule}}

\def\1/2{\hbox{$ {1 \over 2}$ }}
\def\tr{\hbox{Tr}}

\def\h{\hbar}
\def\i/h{{i \over \h}}

\def\inf{\infty}
\def\pd{\partial} 
\def\v{\vec}

\def\a{\alpha} 
\def\b{\beta} 
 
\def\g{\gamma}  
\def\d{\delta} \def\D{\Delta}
\def\l{\lambda} 
\def\e{\epsilon} \def\E{\hbox{$\cal E $}}

\def\s{\sigma}
\def\r{\rho} \def\vr{\varrho}

\def\c{\chi} 
\def\vf{\varphi}

\def\p{\psi}

\def\m{\mu}

\def\tt{\theta}

\def\<{\langle}
\def\>{\rangle}

\def\({\left(}
\def\[{\left[}
\def\){\right)}
\def\]{\right]}
\def\wt{\widetilde}

\usepackage{subfigure}

\usepackage[notcite,notref,color]{showkeys}  

\usepackage{tikz}
\usetikzlibrary{decorations.pathreplacing}
\usetikzlibrary{decorations.pathmorphing}    
\usepackage{multirow}
\usepackage{dcolumn}		
\newcolumntype{.}{D{.}{.}{-1}}
\newcolumntype{i}[1]{D{.}{.}{#1}}

\newcommand{\myfrac}[2]{{\ifmmode{}^{#1}\!/_{\!#2}\else${}^{#1}\!/_{\!#2}$\fi}}

\begin{document}
\sloppy

\title{Ferromagnetic phase in  graphene-based planar heterostructures induced by charged impurity}

\author{P.~Grashin}
\email{grashin.petr@physics.msu.ru} \affiliation{Department of Physics and
Institute of Theoretical Problems of MicroWorld, Moscow State
University, 119991, Leninsky Gory, Moscow, Russia}

\author{K.~Sveshnikov}
\email{costa@bog.msu.ru} \affiliation{Department of Physics and
Institute of Theoretical Problems of MicroWorld, Moscow State
University, 119991, Leninsky Gory, Moscow, Russia}

\date{\today}


\begin{abstract}
The spontaneous self-consistent   generation of axial current and corresponding dipole-like magnetic field in a  planar electron-positron system similar to graphene and related hetero-structures doped by charged impurity with charge $Z$ is explored. It is shown that this effect takes place for $Z \geq Z^{\ast}$ with $Z^{\ast}$ being a peculiar analogue of the Curie point in ferromagnetics. The properties of induced ferromagnetic state are studied in detail. It is shown also that the arising  this way magnetic dipole  leads to a significant decrease of the total Casimir (vacuum) energy of the system, which in turn provides its spontaneous generation above the "Curie point" $Z \geq Z^{\ast}$.
\end{abstract}


\maketitle

\section{Introduction}

A remarkable  feature of two-dimensional crystals like graphene and related planar hetero-structures is that the behavior of charge carriers in such systems is subject to  2+1-dimensional Dirac equation (DE) for  massless (or massive, when it concerns graphene on a substrate) fermions, where the speed of light is replaced by the Fermi velocity $v_F$. The latter circumstance provides a sufficiently larger value of the effective fine structure constant in graphene  compared to that of the 3+1-dimensional QED.  Thereby, in such systems the direct experimental observation of certain essentially non-perturbative QED-effects becomes possible. Between them, one of the most interesting is a deep reconstruction of the vacuum state, caused by discrete levels diving into the lower continuum under influence of a strong quasi-static EM-source.

In the case of graphene it is the charged impurity, which can play the role of such source. The graphene itself serves as the vacuum, while the charge carriers --- electrons and holes --- imitate the  virtual particles. For point-like and extended  Coulomb sources this problem has been intensively studied during two last decades (see, e.g., Refs.~\cite{Katsnelson2006b,*Shytov2007,*Kotov2008, *Pereira2008,*Nishida2014,*Bordag2016,*Bordag2017,*Khalilov2017} and citations therein) and  by taking account of possible screening effects in Refs.~\cite{Voronina2019a,Voronina2019b}.

Of separate interest are the polarization effects in planar QED-systems in presence of an external magnetic field.   In Ref.~\cite{Gornicki1990} there was found the exact solution for the electron motion in the vicinity of a thin tube of magnetic flux. The induced current density is evaluated via direct summation over sea electrons.  For massless fermions the behavior of  induced current, generated by such tube,   was studied in Ref.~\cite{Milstein2011}. In addition, in Ref.~\cite{Gornicki1990} the expression for the vacuum energy density was obtained within the Schwinger-Fock proper time approach, while in Ref.~\cite{Bordag1999} there was considered the evaluation of ground state energy of a spinor field in the background  of a finite radius flux tube with a homogeneous magnetic field inside, based on $\zeta$-function regularization.

Induced magnetic effects  have been also actively studied for the external vector-potentials   of the Aharonov-Bohm  (AB) type, generated by infinitely thin solenoid. In Ref.~\cite{Jackiw2009} the induced charge and current densities are explored for large separations from the solenoid axis. In Refs.~\cite{Khalilov2012,*Khalilov2013} for the crossed point-like Coulomb and magnetic AB potentials  the properties of virtual bound states are considered. In Ref.~\cite{Khalilov2014} by means of  Wichmann-Kroll (WK) contour integration techniques~\cite{Wichmann1956} the analytic expressions for the induced charge and current densities are obtained and their behavior was studied in the limit of small and large distances from the  AB-potential source. The renormalization group analysis of graphene with a supercritical Coulomb impurity combined with magnetic AB potential has been explored in Ref.~\cite{Nishida2016}.

However, so far the polarization effects in planar electron-positron systems haven't been considered for the magnetic fields of dipole type. At the same time, such magnetic fields are of special interest  as physically the most realistic.  The main difficulties, which appear by solving such problems, are caused by absence of analytic solutions of DE for crossed Coulomb-like and  dipole-like magnetic fields. For the planar DE there are known only single  results, which concern either point-like Coulomb and string-like AB sources or a quite different background of vector and scalar Cornell potentials combined with an external magnetic field~\cite{Khalilov2012,Hassanabadi2013}. This circumstance makes impossible the evaluation of the  Casimir (vacuum) energy via representing it as the sum of the integral of the total elastic scattering  phase  and the contribution of discrete spectrum, which is shown to be very powerful in purely Coulomb problems~\cite{Davydov2017, *Voronina2017, *Sveshnikov2017,  Davydov2018b,  Sveshnikov2019b, Voronina2019b}. However,  elaborated recently in Refs.~\cite{Voronina2019c, *Voronina2019d} method of vacuum energy evaluation via logarithmic derivative of the Wronskian solves these problems quite effectively.

In Refs.~\cite{Davydov2018b,  Sveshnikov2019b,Voronina2019b} the behavior of vacuum energy in the strongly coupled planar QED-system similar to graphene in presence of a supercritical extended Coulomb source has been explored. In particular, there was established  the fact of rapid decline of the vacuum  energy in the over-critical region $\sim (-Z^3 / R_0)$ with  $Z>Z_{cr,1}$ and $R_0$ being the charge and radius of the  Coulomb source, which provides complete screening of the electrostatic repulsive self-energy of the external source for  impurity charges  $Z\sim6$ for $\a_g \simeq 0.8$~\cite{Voronina2019b}. Such behavior of the vacuum energy indicates that in the over-critical region with growing impurity charge the induced polarization effects become essentially  non-perturbative.  Therefore, some novel effects could also be expected, including those associated with the magnetic component of  polarization. In particular,  there could take place the effect of spontaneous generation in a self-consistent manner of  the induced axial current and corresponding magnetic field of dipole type. Hereinafter under self-consistent mode  is meant the case, when the induced current produces such magnetic field, which being considered as the external one, generates indeed this vacuum current.

In the present paper this effect is studied for a planar electron-positron system similar to graphene on a substrate under influence of a charged impurity, which creates the potential of a uniformly charged sphere with radius  $R_0$ and exponential decay for $r>R_0$ with the cutoff coefficient $\s$:
\beq
\label{2.1}
A_0(\v{r})=Z|e|\({1 \over R_0}\tt(R_0-r)+{\mathrm{e}^{-\s(r-R_0)} \over r}\tt(r-R_0)\) \ .
\eeq

 The effective fine structure constant is defined as
\beq
\a=e^2/(\h\, v_F\, \e_{eff}) \ , \quad \e_{eff}=(\e+1)/2 \ ,
\eeq
where $\e$ is the dielectric  constant of the substrate, while $v_F=3 t a/2 \h$ is the Fermi velocity in graphene. In the latter expression  $a\simeq \, 1.42\, A\,$ is the distance between nearest carbon atoms in the graphene lattice, while  $t$ is the hopping parameter (the overlap integral between wavefunctions of nearest neighbors in the lattice) and $\l_c=\h/mv_F$ is the effective Compton wavelength \cite{Goerbig2011,*Wallbank2015}.  Here  $m$ denotes the effective fermion mass, which is related to the local energy mismatch in the tight-binding approximation through the relation $\D=2mv_F^2$. These definitions lead to  relation $\l_c/a \simeq 3t/\D$. In this work we restrict to graphene on the SiC substrate with $\a=0.4$ \cite{Pereira2008}   and on the h-BN substrate with $\a=0.8$ \cite{Goerbig2011, Sadeghi2015}. The screening parameter of the external Coulomb field in (\ref{2.1}) is chosen as $\s=1$, while the size of the external Coulomb source (charged impurity) is $R_0=a$. Such cutoff of the Coulomb field at small distances has been introduced earlier in Refs.~\cite{Pereira2008}. For the  SiC substrate the source size $R_0=2a$ is also considered.

Henceforth the system of units in which $\hbar=v_F=m=1$ is used, and so the distances are measured in units of $\l_c$, while the energy --- in units of $mv_F^2$.   For the  SiC substrate the local energy mismatch is $\D=0.26$ eV and therefore   $R_0=1/30$ and $R_0=1/15$ in the  units chosen (corresponding to impurity sizes $a$ and $2a$), while for h-BN one has $\D=0.056$ eV and so $R_0=1/175$.

The self-consistent solution for the magnetic component is sought by means of  successive iterations.  In the first step an external seed current is introduced, whereupon by means of WK techniques the induced current is found. Thereafter the vector-potential, corresponding to this induced current, is calculated and compared with the seed one. The seed current is modified by successive iterations in order to match the seed and induced vector-potentials. In what follows each of the stages of this procedure is described in detail.

The initial seed current  is chosen in the following form
\beq
\label{2.2}
\v{j}\(\v{r}\)=e\,j_0\, \r\, \mathrm{e}^{-\l\sqrt{\r}}\, \d(z)\, \v{e}_{\vf} \ .
\eeq
Henceforth $\v{\r}=\left. \v{r} \right|_{z=0}$ is the planar vector, $\r=|\v{\r}|$. The main purpose for such a choice of the seed current shape is that the main contribution to the vacuum polarization appears from the discrete levels reaching the threshold of the lower continuum, while the dependence $\sim \r\, \mathrm{e}^{-\l\sqrt{\r}}$ is the specific feature of their Dirac currents, built from the  wavefunctions taken directly at the threshold (see e.g., Refs.~\cite{Greiner1985a,Greiner2012}). In the latter case  $\l=2\, \sqrt{8 Z \a}$, whereas in  general $\l$ is a quantity, which should be determined via minimizing the total vacuum energy.

The magnitude of the current $j_0$ can be easily expressed via its dipole moment $\m$ (both are considered in units of $e$)
\begin{equation} \label{2.3}
j_0={\m\, \l^8 \over 10080\,\pi} \ , \quad \m = {1 \over 2\,e}\, \int_V \[\v{r} \times \v{j}\(\v{r}\)\]_z d^3\v{r}  \ .
\end{equation}

Since we are dealing with the electron-positron system on the  plane $z=0$, embedded into the 3-dimensional space, the vector-potential, corresponding to this current, should be evaluated by means of 3-dimensional electrodynamics:
\beq
\label{2.4}
\v{A}\(\v{r}\)=\int_V\, d^3\v{r}^\prime\, {\v{j}\(\v{r}^\prime\) \over |\v{r}-\v{r}^\prime|} \ ,
\eeq
that gives
\beq
\label{2.5}
\bal
&\v{A}\(\v{r}\)|_{z=0}=\v{e}_\vf\, A_\vf(\r) \ , \\ & A_\vf(\r)={4ej_0}\, \Bigg[\int_0^\r d\r^\prime\, \r^{\prime}\,\mathrm{e}^{-\l\sqrt{\r^\prime}}\,\(K\(\r^{\prime2} \over \r^2\)-E\(\r^{\prime2} \over \r^2\)\)  \\& + \ {1 \over \r}\int_\r^\inf d\r^\prime \, \r^{\prime2}\, \mathrm{e}^{-\l\sqrt{\r^\prime}}\(K\(\r^2 \over \r^{\prime2}\)-E\(\r^2 \over \r^{\prime2}\)\)\Bigg] \ ,
\eal
\eeq
with  $K(z)$ and $E(z)$ being the complete elliptic integrals. The asymptotics $A_\vf(\r\to 0) \to e \m \l^6 \r/ 5040 + O(\r^2)$ and $A_\vf(\r\to \inf) \to e \m /\r^2 + O(1/\r^3)$
provide that the seed magnetic field describes a spatially distributed dipole-like configuration.

\section{Evaluation of the induced charge and current via WK  techniques}

The starting points for WK approach  are the following expressions for the induced charge and current densities~\cite{Wichmann1956,Gyulassy1975,Greiner1985a,Greiner2012}
\begin{multline} \label{4.5a}
\vr_{vac}(\vec{\r})=-\frac{|e|}{2}\(\sum\limits_{\e_{n}<\e_{F}} \p_{n}(\vec{\r})^{\dagger}\p_{n}(\vec{\r}) \ - \right. \\ \left. - \ \sum\limits_{\e_{n}\geqslant \e_{F}} \p_{n}(\vec{\r})^{\dagger}\p_{n}(\vec{\r}) \) \ ,
\end{multline}
and
\begin{multline} \label{4.5b}
\v{j}_{vac}(\vec{\r})=-\frac{|e|}{2}\(\sum\limits_{\e_{n}<\e_{F}} \p_{n}(\vec{\r})^{\dagger}\v{\a}\,\p_{n}(\vec{\r}) \ - \right. \\ - \  \left. \sum\limits_{\e_{n}\geqslant \e_{F}} \p_{n}(\vec{\r})^{\dagger}\v{\a}\,\p_{n}(\vec{\r}) \) \ ,
\end{multline}
where $\e_F$ is the Fermi level, which in such problems with strong Coulomb fields should be chosen at the lower threshold ($\e_F=-1$), while $\e_{n}$ and $\p_n(\vec{\r})$ represent the eigenvalues and eigenfunctions of  DE.

The spectral DE takes the form
\beq
\label{4.1}
\[\v{\a}\,(\v{p}-e\v{A}\(\v{\r}\))+\b+V(\r)\]\p=\e\,\p \ ,
\eeq
where $V(\r)=e A_0(\r)$ and the 4-dimensional standard representation of Dirac matrices, where $\b$ is diagonal, is used.

Since we are dealing with the axially-symmetric problem, $j_z$ is conserved and so the radial and angle variables are separated via substitution
\beq\label{4.2}
\p_{m_j}\(\v{\r}\)={1 \over \sqrt{2\pi \r}}\left(
\begin{array}{c}
a_{m_j}(\r)\,e^{i(m_j - 1/2)\vf}\\
b_{m_j}(\r)\,e^{i(m_j + 1/2)\vf}\\
-ic_{m_j}(\r)\,e^{i(m_j - 1/2)\vf}\\
-id_{m_j}(\r)\,e^{i(m_j + 1/2)\vf}
\end{array}
\right) \ ,
\eeq
with $m_j$ being the eigenvalue of $j_z$, while the radial functions $a_{m_j}(\r), b_{m_j}(\r), c_{m_j}(\r), d_{m_j}(\r)$ for real-valued $\e$ can be also taken real.

As a result, the initial DE (\ref{4.1}) splits into two independent radial subsystems, which take the form
\beq\begin{gathered}
\label{4.3}
(\pd_\r-{m_j \over \r}-|e|A_\vf(\r))\,a_{m_j}(\r)= (\e+1-V(\r))\,d_{m_j}(\r) \ , \\
(\pd_\r+{m_j \over \r}+|e|A_\vf(\r))\,d_{m_j}(\r)= -(\e-1-V(\r))\,a_{m_j}(\r) \ ,
\end{gathered}
\eeq
and
\beq\begin{gathered}
\label{4.4}
(\pd_\r+{m_j \over \r}+|e|A_\vf(\r))\,b_{m_j}(\r)= (\e+1-V(\r))\,c_{m_j}(\r) \ , \\
(\pd_\r-{m_j \over \r}-|e|A_\vf(\r))\,c_{m_j}(\r)= -(\e-1-V(\r))\,b_{m_j}(\r) \ .
\end{gathered}
\eeq
For what follows it is important that the subsystems (\ref{4.3}) and (\ref{4.4}) are related via replacement: $a_{m_j}(\r)\to b_{m_j}(\r), \, d_{m_j}(\r)\to c_{m_j}(\r), \, m_j\to -m_j, \, A_{\vf}\to - A_{\vf}$ and vice versa.

The essence of the WK techniques is the representation of the densities (\ref{4.5a}-\ref{4.5b}) in terms of contour integrals on the first sheet of the Riemann energy plane containing the trace of the Green function of DE. The formal expression for the Green function reads
\beq
\label{4.7}
G(\vec{\r},\vec{\r}\ ';\e)=\sum\limits_{n}\frac{\p_{n}(\vec{\r})\,\p_{n}(\vec{\r}\ ')^{\dagger}}{\e_{n}-\e} \ .
\eeq
\begin{figure}
\center
\includegraphics[scale=0.2]{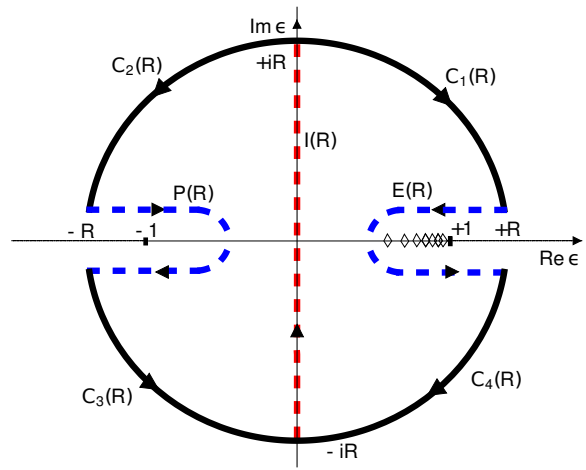}
\caption{\small WK contours in the complex energy plane, used for representation of the induced charge and current densities via  contour integrals. The direction of contour integration is chosen in correspondence with  (\ref{4.7}).}
\label{pic:1}
\end{figure}
By means of  the Green function trace the induced densities (\ref{4.5a}-\ref{4.5b}) can be rewritten as the following integrals along the WK contours $P(R)$ and $E(R)$, shown in Fig.\ref{pic:1},
\beq\label{4.8a}
\vr_{vac}(\vec{\r})=-\frac{|e|}{4 \pi i} \lim_{R\to\inf}\(\int\limits_{P(R)}+\int\limits_{E(R)}\) d\e\, \mathrm{Tr} \, G(\vec{\r},\vec{\r};\e) \ ,
\eeq
\beq
\label{4.8b}
\v{j}_{vac}(\vec{\r})=-\frac{|e|}{4 \pi i} \lim_{R\to\inf}\(\int\limits_{P(R)}+\int\limits_{E(R)}\) d\e\, \mathrm{Tr}\[\v{\a} \, G(\vec{\r},\vec{\r};\e)\]  \ .
\eeq

In the next step, due to axial symmetry of the problem one obtains for $\mathrm{Tr} \, G(\vec{\r},\vec{\r};\e)$ and $\mathrm{Tr} \[\v{\a} \, G(\vec{\r},\vec{\r};\e)\]$  the following representation in terms of partial series in $m_j$
\beq
\label{4.9}
\bal
&\mathrm{Tr} \, G(\vec{\r},\vec{\r};\e)={1 \over 2\pi \r}\sum_{m_j=\pm1/2,3/2,...}\mathrm{Tr} \, G_{m_j}(\r,\r;\e)\, ,
\\&\mathrm{Tr}\[\v{\a} \, G(\vec{\r},\vec{\r};\e)\]={\v{e}_{\vf} \over 2\pi \r }\sum_{m_j=\pm1/2,3/2,...}\mathrm{Tr}\[\a_\vf \, G_{m_j}(\r,\r;\e)\]\, ,
\eal
\eeq
where $G_{m_j}(\r,\r';\e)$ is the sum of radial Green functions for the spectral problems (\ref{4.3}-\ref{4.4}), which in turn are built from the regular for $\r \to 0$ or $\r \to +\infty$ solutions of systems (\ref{4.3})-(\ref{4.4}). Denoting the regular for $\r \to 0$ solutions by label $0$, while the regular for $\r \to +\infty$ ones by $\infty$, the expressions for $\mathrm{Tr} \, G_{m_j} (\r,\r;\e)$ and $\mathrm{Tr}\[\a_\vf \, G_{m_j}(\r,\r;\e)\]$ can be written as

\beq
\label{4.10}
\begin{aligned}
	&\mathrm{Tr}\, G_{m_j}(\r,\r;\e)=\\&{1 \over J_{m_j}^{ad}(\e)}\(a_{m_j}^0(\r)a_{m_j}^\infty(\r)+d_{m_j}^0(\r)d_{m_j}^\infty(\r)\) + \\ & + {1 \over J_{m_j}^{bc}(\e)}\(b_{m_j}^0(\r)b_{m_j}^\infty(r)+c_{m_j}^0(\r)c_{m_j}^\infty(\r)\) \ ,
\end{aligned}
\eeq

\beq
\label{4.11}
\begin{aligned}
	&\mathrm{Tr}\[ \a_{\vf} G_{m_j}(\r,\r;\e)\]=\\&-{1 \over J_{m_j}^{ad}(\e)}\(d_{m_j}^0(\r)a_{m_j}^\infty(\r)+a_{m_j}^0(r)d_{m_j}^\infty(\r)\) + \\ & + {1 \over J_{m_j}^{bc}(\e)}\(c_{m_j}^0(\r)b_{m_j}^\infty(\r)+b_{m_j}^0(\r)c_{m_j}^\infty(\r)\) \ ,
\end{aligned}
\eeq
with $J^s_{m_j}(\e)$ being the Wronskians of systems (\ref{4.3}-\ref{4.4}), namely
\beq\begin{gathered}
\label{4.12}
J_{m_j}^{ad}(\e)=\(d^0(\r)a^\infty(\r)-a^0(\r)d^\infty(\r)\) \ , \\
J_{m_j}^{bc}(\e)=\(c^0(\r)b^\infty(\r)-b^0(\r)c^\infty(\r)\) \ ,
\end{gathered}\eeq
which provide the correct normalization of $G_{m_j}(\r,\r';\e)$.

Proceeding further, upon  deformation of the WK contours $P(R)$ and $E(R)$  to  imaginary axis (see Fig.\ref{pic:1}) the final expressions for the vacuum charge and current densities take the form
\beq
\label{4.13a}
\vr_{vac}(\v{\r})=\sum_{m_j=1/2,\, 3/2,...}\vr_{vac,|m_j|}(\r) \ ,
\eeq
where
\beq\label{4.13}
\vr_{vac,|m_j|}(\r)={|e| \over \(2\pi\)^2\r}\int_{-\infty}^\infty dy \ \mathrm{Re}\[\mathrm{Tr}\, G_{|m_j|}(\r,\r;iy)\] \ ,
\eeq
\beq
\label{4.14}
\mathrm{Tr}\, G_{|m_j|}(\r,\r;iy)=\mathrm{Tr}\, G_{m_j}(\r,\r;iy)+\mathrm{Tr}\, G_{-m_j}(\r,\r;iy) \ ,
\eeq
and
\beq
\label{4.15}
\v{j}_{vac}(\v{\r})=\v{e}_\vf\,\sum_{m_j=1/2,\, 3/2,...}j_{vac,|m_j|}(\r) \ ,
\eeq
where
\beq
\label{4.16}
j_{vac,|m_j|}(\r)={|e| \over \(2\pi\)^2\r}\int_{-\infty}^\infty dy \ \mathrm{Re}\[\mathrm{Tr}\[ \a_{\vf} G_{|m_j|}(\r,\r;iy)\]\] \ ,
\eeq
\beq\begin{gathered}
\label{4.17}
\mathrm{Tr}\[\a_{\vf}G_{|m_j|}(\r,\r;iy)\]=\mathrm{Tr}\[ \a_{\vf}G_{m_j}(\r,\r;iy)\] \ + \\ + \ \mathrm{Tr}\[ \a_{\vf}G_{-m_j}(\r,\r;iy)\] \ .
\end{gathered}\eeq
In presence of negative discrete levels with $-1\leqslant \e_n<0$ these formulae transform into~\cite{Gyulassy1975}
\begin{multline}
\label{4.18}
\vr_{vac,|m_j|}(\r)= \\ {|e| \over 2\pi}\[\sum_{m_j= \pm|m_j|}\sum_{-1\leq\e_n<0}\c_{n,m_j}(\r)^T\c_{n,m_j}(\r)\ + \right.\\ \left. + \ {1 \over 2\pi \r}\int_{-\infty}^\infty dy \ \mathrm{Re}\[\mathrm{Tr} G_{|m_j|}(\r,\r;iy)\]\] \ ,
\end{multline}
\begin{multline}
\label{4.19}
j_{vac,|m_j|}(\r)= \\ {|e| \over 2\pi}\[\sum_{m_j= \pm|m_j|}\sum_{-1\leq\e_n<0}\c_{n,m_j}(\r)^T  A \c_{n,m_j}(\r)\ + \right.\\ \left. + \ {1 \over 2\pi \r}\int_{-\infty}^\infty dy \ \mathrm{Re}\[\mathrm{Tr}\[ \a_{\vf} G_{|m_j|}(\r,\r;iy)\] \] \]\ \ ,
\end{multline}
where
\beq\label{4.19a}
A=\begin{pmatrix}0&0&0&-1\\
	0&0&1&0\\
	0&1&0&0\\
	-1&0&0&0\\
\end{pmatrix} \ ,
\eeq
while $\c_{n,m_j}(\r)$ is the real-valued radial Dirac wavefunction of the eigenstate with $-1\leq\e_n<0$, consisting of components $a_{n,m_j}(\r), b_{n,m_j}(\r), c_{n,m_j}(\r), d_{n,m_j}(\r)$.

It should be noted that by derivation of relations (\ref{4.13})-(\ref{4.19}) the following general properties of the Green functions under complex conjugation
\beq
\label{4.20}
G_{m_j}(Q,\m;\r,\r;\e)^*=G_{m_j}(Q,\m;\r,\r;\e^*)\, ,
\eeq
as well as the properties of their traces under changing the sign of external fields ($Q\to-Q$, $\m\to-\m$)
\begin{multline}
\label{4.21}
\tr G_{m_j}(Q,\m;\r,\r;\e)=-\tr G_{m_j}(-Q,\m;\r,\r;-\e) \ , \\ \tr\[ \a_\vf G_{-m_j}(Q,\m;\r,\r;\e)\]= \\ - \tr\[ \a_\vf G_{m_j}(Q,-\m;\r,\r;\e)\] \ ,
\end{multline}
play an essential role.
Namely, due to these properties  there follow the relations
\begin{multline}
\label{4.22}
\tr G_{m_j} (Q,\m;\r,\r; i y)^{\ast}=-\tr G_{m_j}(-Q,\m;\r,\r;i y) \ , \\ \tr\[ \a_\vf G_{m_j}(Q,\m;\r,\r;iy)\]^*= \\ \tr\[ \a_\vf G_{-m_j}(-Q,-\m;\r,\r;iy)\] \ .
\end{multline}
At the same time, on account of the Furry theorem the induced vacuum charge and current densities should be odd functions of external EM-fields, whence it follows that both $\vr_{vac,|m_j|}(\r)$ and $j_{vac,|m_j|}(\r)$ are defined only by $\mathrm{Re}\[\mathrm{Tr}\, G_{|m_j|}(\r,\r;iy)\]$ and $\mathrm{Re}\[\mathrm{Tr}\[ \a_{\vf} G_{|m_j|}(\r,\r;iy)\]\]$,  which has already been used  in expressions (\ref{4.13})-(\ref{4.19}).

Since the induced charge and current densities are represented as infinite series in $m_j$, there appears the question of their convergence. The solution of this problem is closely related to the renormalization of these quantities. For the planar case  the renormalization of the induced charge density in absence of  magnetic field has been considered in detail in Refs.~\cite{Davydov2018a, Sveshnikov2019a, Voronina2019a}.  In the present case with the dipole-like magnetic field the crucial role is played by the fact that  the corresponding vector-potential   behaves for $\r \to 0$ and $\r \to \infty$ as $A_\vf \sim \r$ and $A_\vf \sim 1/\r^2$, respectively. So the main properties of the regular for $\r \to 0$ or $\r \to +\infty$ solutions of systems (\ref{4.3}-\ref{4.4}) remain unchanged with respect to the purely Coulomb case, and hence, the asymptotics of $\tr G_{m_j}$ for $y\to\infty$ also does not change. More concretely,  $y \to \inf$ is equivalent to the high-energy limit of DE, where the localized smooth magnetic dipole shows up just as a perturbation in the Coulomb field background. This result justifies also the  WK contours deformation to imaginary axis, considered above, because in the purely Coulomb planar systems the latter  is reliably verified in Refs.~\cite{Davydov2018a, Sveshnikov2019a, Voronina2019a}.  In turn, the properties of the limit $|m_j| \to \inf$  follow directly from the systems (\ref{4.3}-\ref{4.4}), since in this case the magnetic field again turns out to be a perturbation in the Coulomb background. Remark that these considerations are invalid in the AB case, when $A_\vf \sim 1/\r$ for $\r \to +\infty$.

Due to these arguments there follows that in this case the renormalization of the induced charge density should be based on the same procedure as in absence of  magnetic field. Namely, first  $\vr_{vac,|m_j|}^{(3+)}(\r)$ is introduced via relation
\begin{multline}
\label{4.24}
\vr_{vac,|m_j|}^{(3+)}(\r)= \\ {|e| \over 2\pi}\,\Bigg[\sum_{m_j=\pm|m_j|}\sum_{-1\leq\e_n<0}\c_{n,m_j}(\r)^T \c_{n,m_j}(\r) \ + \\  {1 \over \pi \r}\int_0^\infty dy\,\mathrm{Re}\[\mathrm{Tr}\, G_{|m_j|}(\r,\r;iy)-\mathrm{Tr}\, G_{|m_j|}^{(1)}\(\r;iy\)\]\Bigg] \ .
\end{multline}
By construction  $\vr_{vac,|m_j|}^{(3+)}(\r)$ do not contain linear in the external Coulomb field terms. The latter is provided by the fact that $G_{|m_j|}^{(1)}\(\r;iy\)$ is the linear in $Q$ component of the partial Green function $G_{|m_j|}(\r,\r;iy)$, defined  as
\beq
\label{4.25}
G_{|m_j|}^{(1)}=Q\, \left.\({\pd G_{|m_j|} \over \pd Q}\)\right|_{Q=0} \ .
\eeq
The explicit answer for $\mathrm{Re}\[\mathrm{Tr}\, G_{|m_j|}^{(1)}\(\r;iy\)\]$, found via first Born approximation for $G_{m_j}$, reads
\begin{multline}
\label{4.26}
\mathrm{Re}\[\mathrm{Tr}\, G_{|m_j|}^{(1)}\(\r;iy\)\]=\\ - \r\,\Bigg[2K_{|m_j-1/2|}^2(\wt{\g}\r)\int_0^\r d\r^\prime \,\r^\prime V(\r^\prime)\(\(1-y^2\)I_{|m_j-1/2|}^2(\wt{\g}\r^\prime)\  \right. \\ \left. + \ \(1+y^2\)I_{|m_j+1/2|}^2(\wt{\g}\r^\prime)\) \ +  \\  2K_{|m_j+1/2|}^2(\wt{\g}\r)\int_0^\r d\r^\prime \,\r^\prime V(\r^\prime)\(\(1-y^2\)I_{|m_j+1/2|}^2(\wt{\g}\r^\prime)  \right. \\ \left. + \ (1+y^2)I_{|m_j-1/2|}^2(\wt{\g}\r^\prime)\) \ +  \\   2I_{|m_j-1/2|}^2(\wt{\g}\r)\int_\r^\infty d\r^\prime \,\r^\prime V(\r^\prime)\(\(1-y^2\)K_{|m_j-1/2|}^2(\wt{\g}\r^\prime)  \right. \\ \left. + \ \(1+y^2\)K_{|m_j+1/2|}^2(\wt{\g}\r^\prime)\) \ +  \\  2I_{|m_j+1/2|}^2(\wt{\g}\r)\int_\r^\infty d\r^\prime \,\r^\prime V(\r^\prime)\(\(1-y^2\)K_{|m_j+1/2|}^2(\wt{\g}\r^\prime)  \right. \\ \left. + \ (1+y^2)K_{|m_j-1/2|}^2(\wt{\g}\r^\prime)\)\Bigg]  \  ,
\end{multline}
where and in that follows
\beq
\label{4.26a}
\wt{\g}=\sqrt{1+y^2} \ .
\eeq

After these preliminary steps the renormalization of the vacuum charge density reduces to replacement of the linear term by renormalized first order perturbative density  $\vr_{vac}^{(1)}(\v{\r})$, which  has been evaluated in Refs.~\cite{Davydov2018a, Sveshnikov2019a, Voronina2019a}, and takes the form
\beq
\label{4.23}
\vr_{vac}^{ren}(\v{\r})=\vr_{vac}^{(1)}(\v{\r})+\sum_{m_j=1/2,\, 3/2,...}\vr_{vac,|m_j|}^{(3+)}(\r) \ .
\eeq

The general properties of the renormalized induced charge density coincide with those in absence of magnetic field  (see, e.g., Refs.~\cite{Davydov2018a,  Voronina2019a, Sveshnikov2019a} and citations therein). In particular, the expression (\ref{4.23}) provides vanishing of the total induced charge  $Q_{vac}^{ren}=\int d^2 \v{\r} \, \vr_{vac}^{ren}(\v{\r})$ for $Z<Z_{cr, 1}$. The term $Q_{vac}^{(1)}=\int d^2\v{\r}\, \vr_{vac}^{(1)}(\v{\r})$ disappears as the first order effect of the perturbation theory, whereas vanishing  of the contribution from $\vr_{vac,|m_j|}^{(3+)}(\r)$ to $Q_{vac}^{ren}$ is verified via direct numerical calculation (in more details this question is considered in Ref.~\cite{Davydov2018a} and especially in Ref.~\cite{Sveshnikov2019a}). The change in $Q_{vac}^{ren}$ can occur only for $Z>Z_{cr,1}$ due to discrete levels diving    into the lower continuum, wherein each dived level $\p_{n,m_j}(\v{\r})$ shifts the total induced charge by $(-|e|)$, while the charge density  $\vr_{vac}^{ren}(\v{\r})$ undergoes the following change \cite{Fano1961,Greiner2012}
\beq\label{4.27a}
\D\vr_{vac}^{ren}(\v{\r})=-|e|\,\p_{\e_n=-1,m_j}(\v{\r})^\dagger \p_{\e_n=-1,m_j}(\v{\r}) \ .
\eeq

Renormalization of the current density proceeds in the same way. Now the linear in $\m$ terms are extracted from the expression for $\mathrm{Tr}\[\a_{\vf}G_{|m_j|}(\r,\r;iy)\]$ and replaced by the renormalized perturbative current density $j_{vac,|m_j|}^{(1)}(\r)$, which is determined quite similar to $\vr_{vac}^{(1)}(\v{\r})$ and in the case of magnetic dipole doesn't vanish only for $|m_j|=1/2\, , 3/2$. For these purposes one finds first the  induced current density component $j_{vac,|m_j|}^{(3+)}(\r)$, which is defined as a complete analogue of $\vr_{vac,|m_j|}^{(3+)}(\r)$
\begin{multline}
\label{4.27}
j_{vac,|m_j|}^{(3+)}(\r)=\\ = {|e| \over 2\pi}\Bigg[\sum_{m_j=\pm|m_j|}\sum_{-1\leq\e_n<0}\c_{n,m_j}(\r)^T A \c_{n,m_j}(\r) \ +  \\ + \ {1 \over \pi \r}\int_0^\infty dy\,\mathrm{Re}\Big(\mathrm{Tr}\[\a_{\vf} G_{|m_j|}(\r,\r;iy)\] \ -  \\ - \ \mathrm{Tr}\[\a_{\vf} G_{|m_j|}^{(1)}\(\r;iy\)\]\Big)\Bigg] \ .
\end{multline}

The explicit answer for $\mathrm{Re}\[\mathrm{Tr}\[ \a_{\vf} G_{|m_j|}^{(1)}\(\r;iy\)\]\]$ reads
\begin{multline}
\label{4.28}
  \mathrm{Re}\[\mathrm{Tr}\[ \a_{\vf} G_{|m_j|}^{(1)}\(\r;iy\)\]\]= \\ = 8|e|\,\wt{\g}\r\, \Big[K_{|m_j-1/2|}(\wt{\g}\r)(K_{|m_j+1/2|}(\wt{\g}\r)  \ \times \\  \int_0^\r d\r^\prime \, \r^\prime A_\vf(\r^\prime)I_{|m_j-1/2|}(\wt{\g}\r^\prime)I_{|m_j+1/2|}(\wt{\g}\r^\prime) \ + \\ + \ I_{|m_j-1/2|}(\wt{\g}\r)I_{|m_j+1/2|}(\wt{\g}\r) \ \times \\ \int_\r^\infty d\r^\prime \, \r^\prime A_\vf(\r^\prime)K_{|m_j-1/2|}(\wt{\g}\r^\prime)K_{|m_j+1/2|}(\wt{\g}\r^\prime) \Big] \ .
\end{multline}

As a result, the renormalized induced current density is given by the following expression
\beq\begin{gathered}
\label{4.29}
\v{j}_{vac}^{ren}(\v{\r})=\v{e}_\vf\, j_{vac,\vf}^{ren}(\r) \ , \\ j_{vac,\vf}^{ren}(\r)=j_{vac}^{(1)}(\r)+\sum_{m_j=1/2,\, 3/2,...}j_{vac,|m_j|}^{(3+)}(\r) \ .
\end{gathered}\eeq
By discrete levels diving into the lower continuum $j_{vac,\vf}^{ren}(\r)$ undergoes jumps, which are described by the corresponding analogue of the  Fano rule (\ref{4.27a})
\beq
\D \v{j}_{vac}^{ren}(\v{\r})=-|e|\,\p_{\e_n=-1,m_j}(\v{\r})^\dagger \v{\a}\, \p_{\e_n=-1,m_j}(\v{\r}) \ .
\eeq
The integral current $J_{vac,\vf}^{ren}=\int_0^\inf \r\, d\r \, j_{vac,\vf}^{ren}(\r)$ passing through the half-axis with fixed axial angle $\vf$ also changes jump-like, but the value of this jump depends on the structure of wavefunction of the dived level and in contrast to $Q_{vac}^{ren}$ is in general  not quantized.

\section{Induced current in graphene}

\begin{figure*}[ht!]
\subfigure[]{
		\includegraphics[scale=0.52]{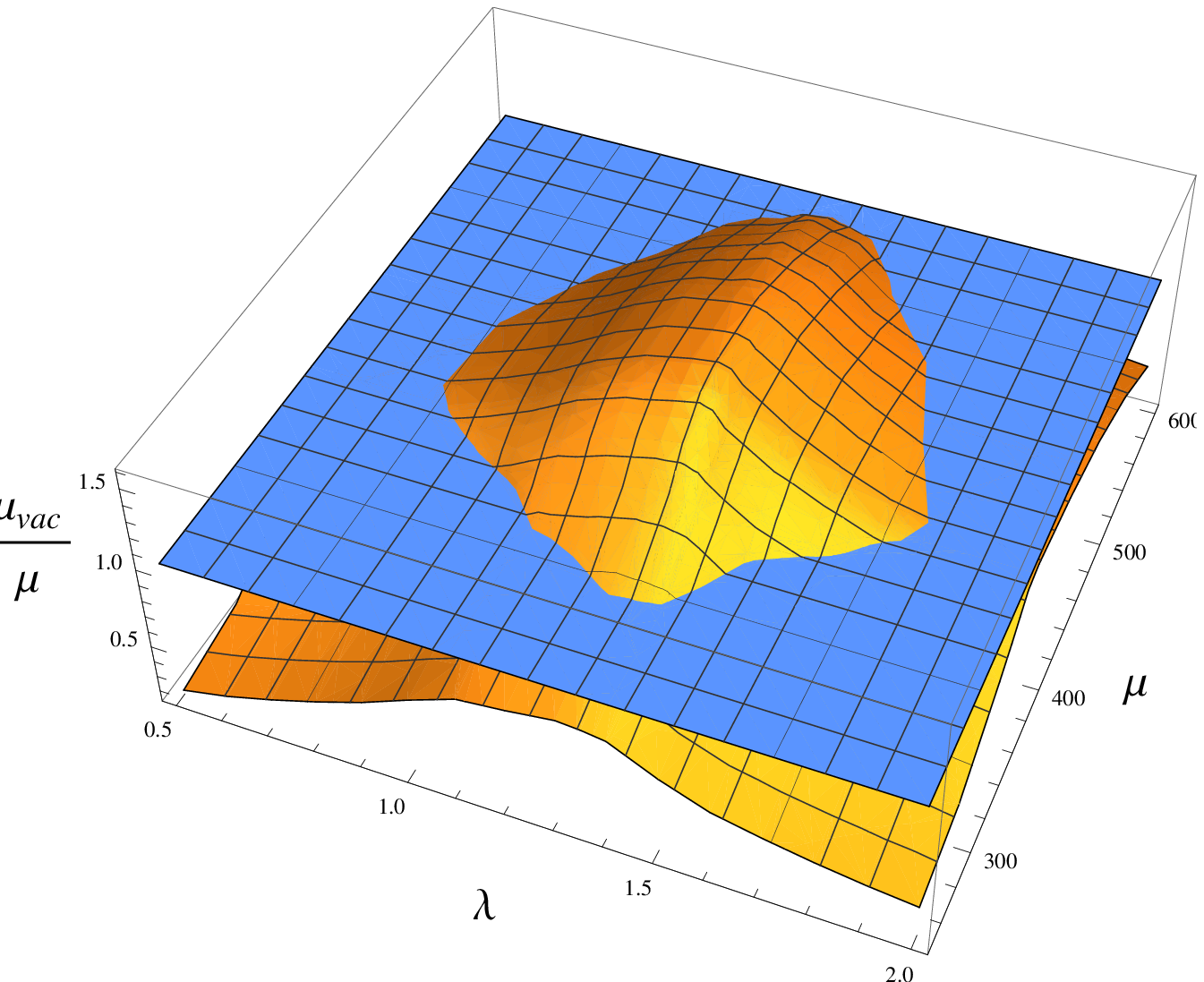}
}
\hfill
\subfigure[]{
		\includegraphics[scale=0.55]{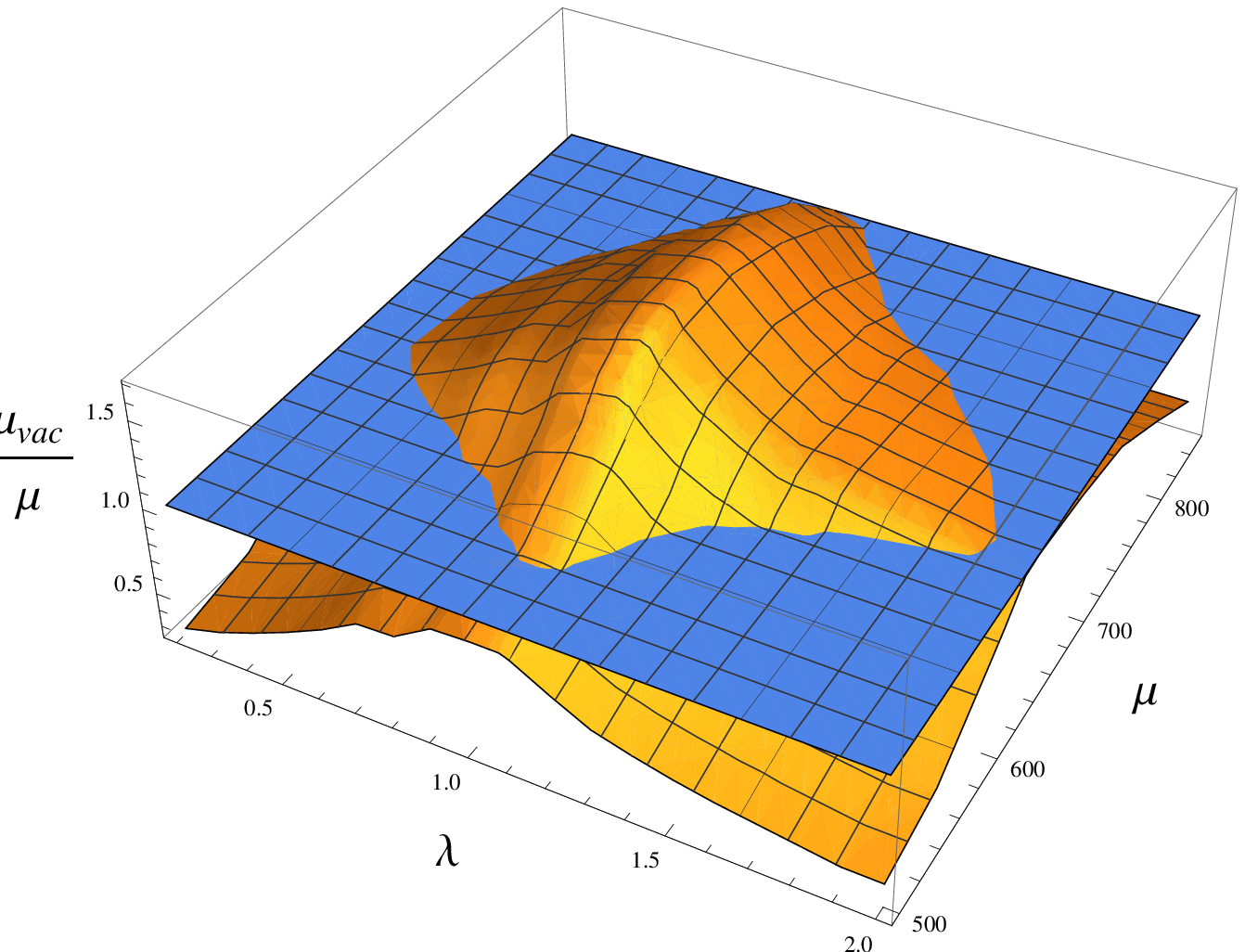}
}
\caption{The ratio between induced and seed magnetic moments $\m_{vac}/\m$ for $Z=10$ and (a) $\a=0.4$ and $R_0=1/30$; (b) $\a=0.8$ and $R_0=1/175$. Closed loops at the intersection of the surface of the evaluated ratio $\m_{vac}/\m$, considered as a function of  the seed current parameters $\l$ and $\m$, with the plane $\m_{vac}/\m=1$, represent those sets of $\l$ and $\m$, which provide   the  self-consistent generation of the induced current.}
	\label{pic2}
\end{figure*}
\begin{figure*}[ht!]
\subfigure[]{
		\includegraphics[scale=0.5]{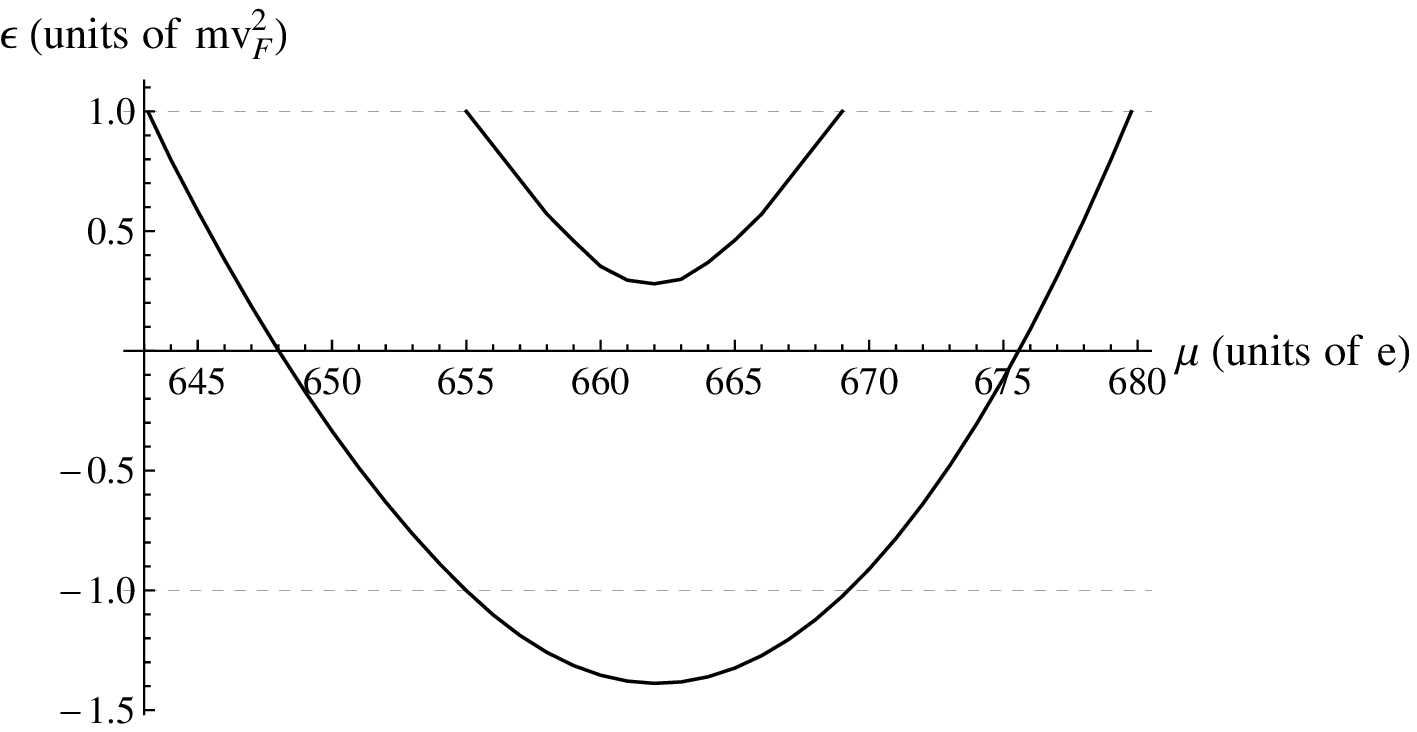}
}
\hfill
\subfigure[]{
		\includegraphics[scale=0.5]{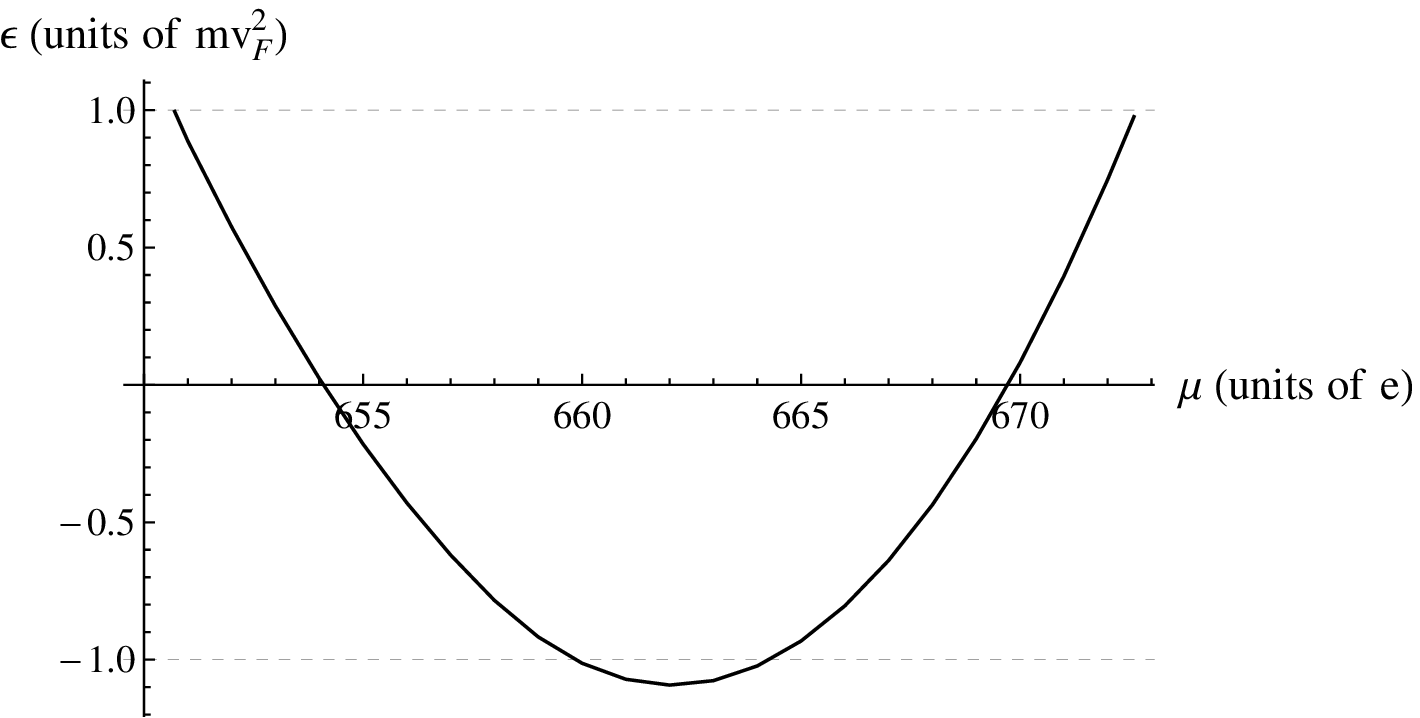}
}
\caption{The trajectories of the lowest discrete levels for  $\m$ varying in the vicinity of the "saddle point" $\m^{\ast}$ for $Z=10$, $\a=0.8$, $R_0=1/175$ and  $\l=1.863$: (a) ad-subsystem, $m_j=1/2$; (b) bc-subsystem, $m_j=-1/2$.}
	\label{pic3}
\end{figure*}
\begin{figure*}[ht!]
\subfigure[]{
		\includegraphics[scale=0.55]{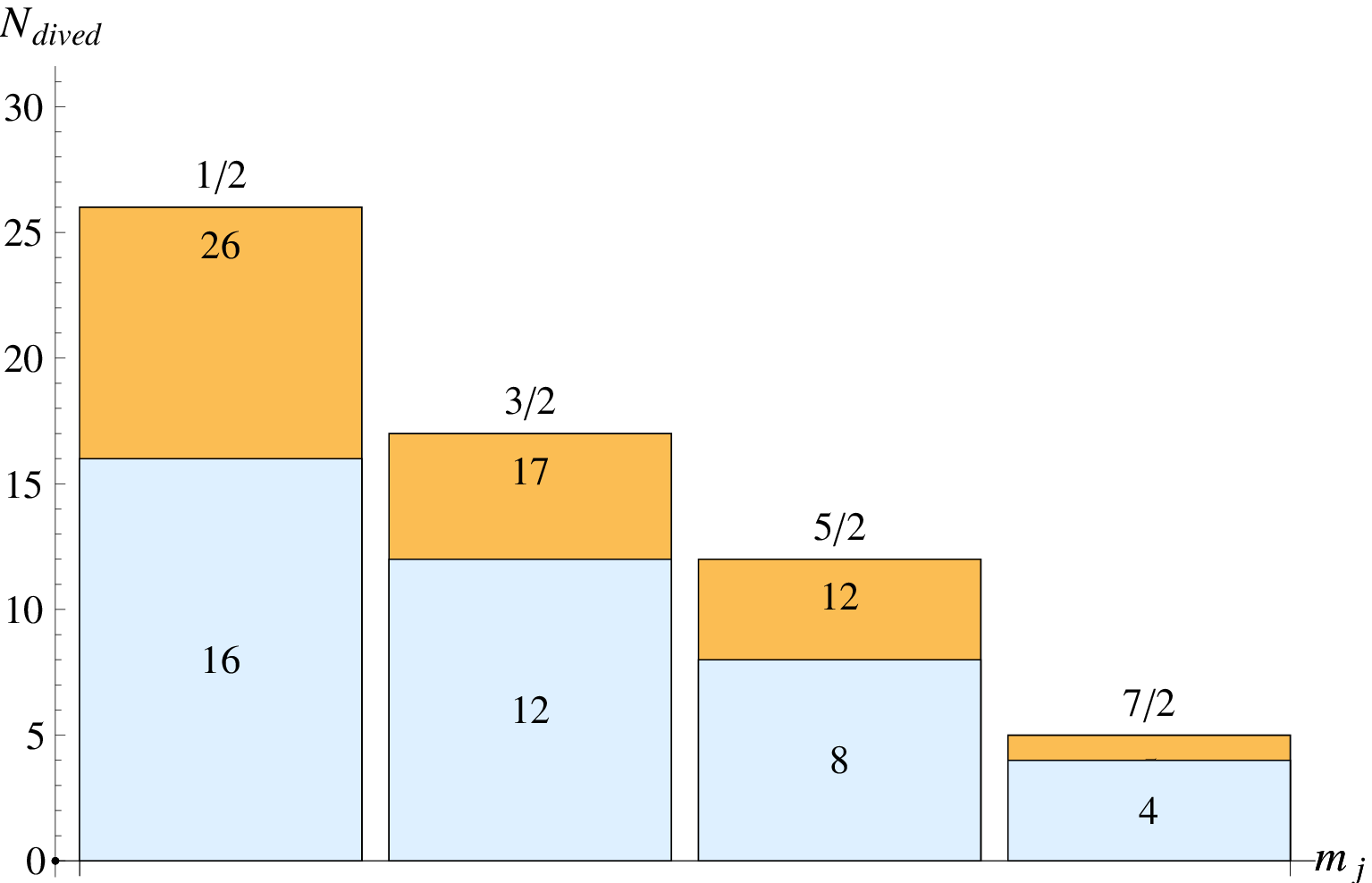}
}
\hfill
\subfigure[]{
		\includegraphics[scale=0.55]{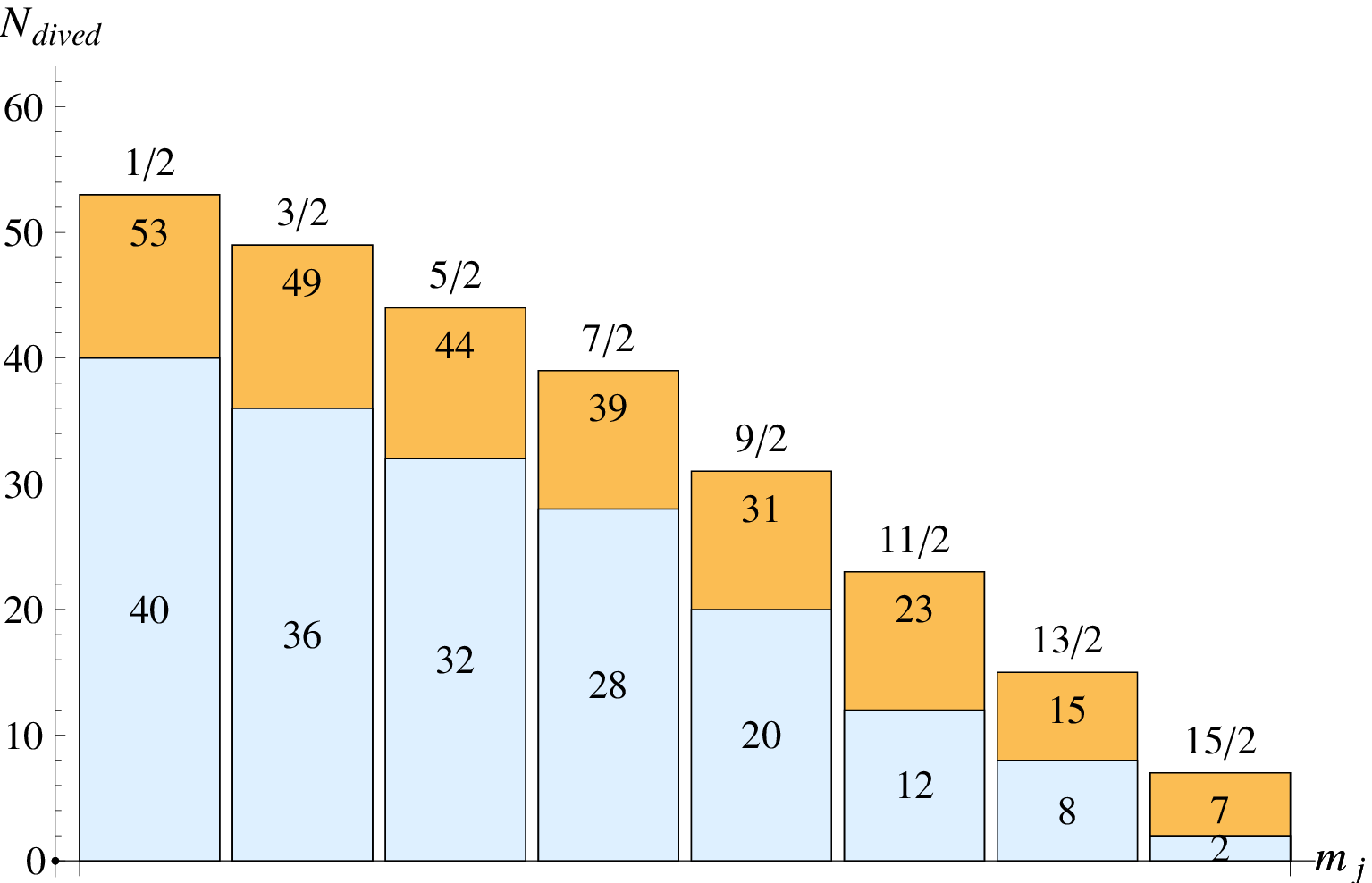}
}
\caption{ The dived discrete levels distribution per partial channel with (orange) and without (turquoise) magnetic field for  $Z=10$ and  (a) $\a=0.4$, $R_0=1/30$, $\l=1.689$ and $\m=373.69$; (b) $\a=0.8$, $R_0=1/175$, $\l=1.863$ and $\m=670.45$.}
	\label{pic4}
\end{figure*}
\begin{figure*}[ht!]
\subfigure[]{
		\includegraphics[scale=0.55]{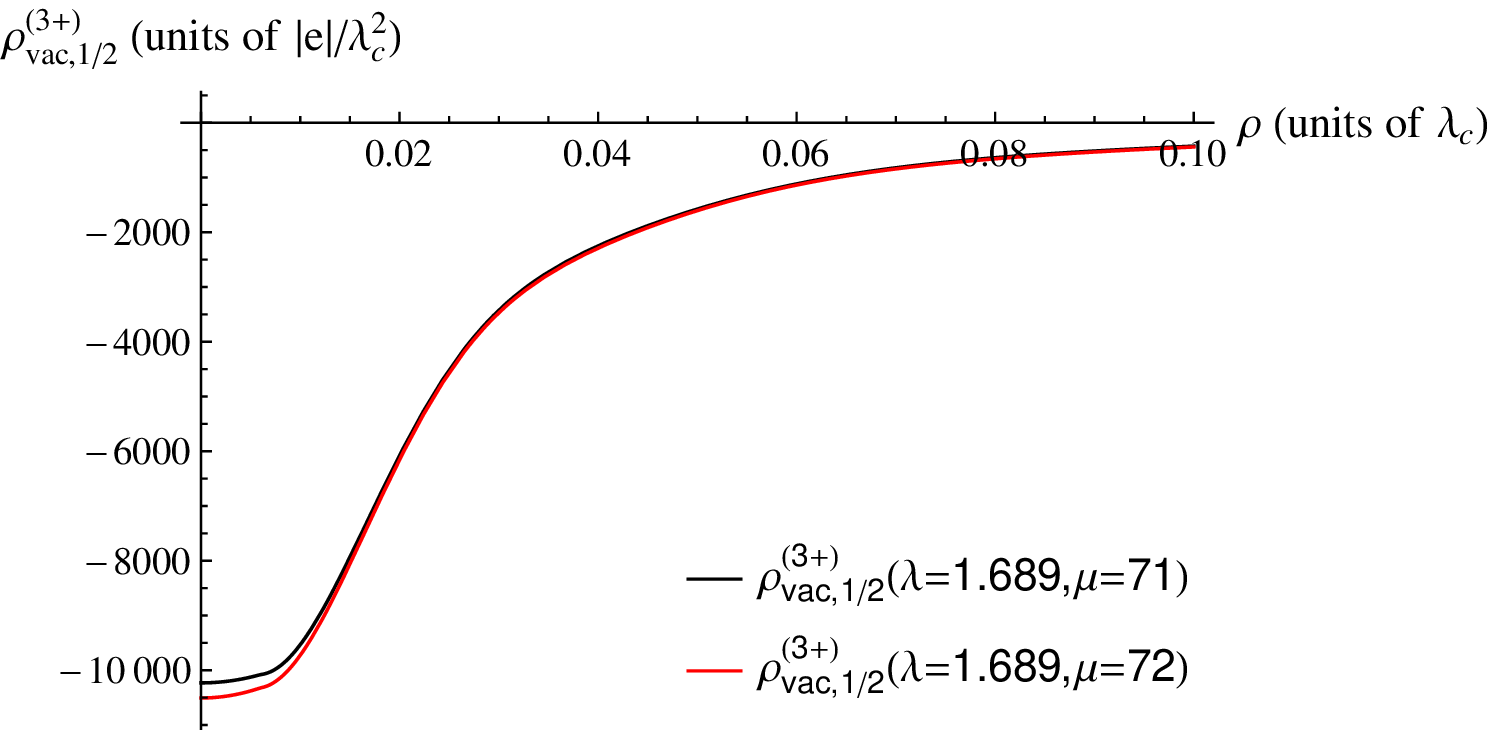}
}
\hfill
\subfigure[]{
		\includegraphics[scale=0.55]{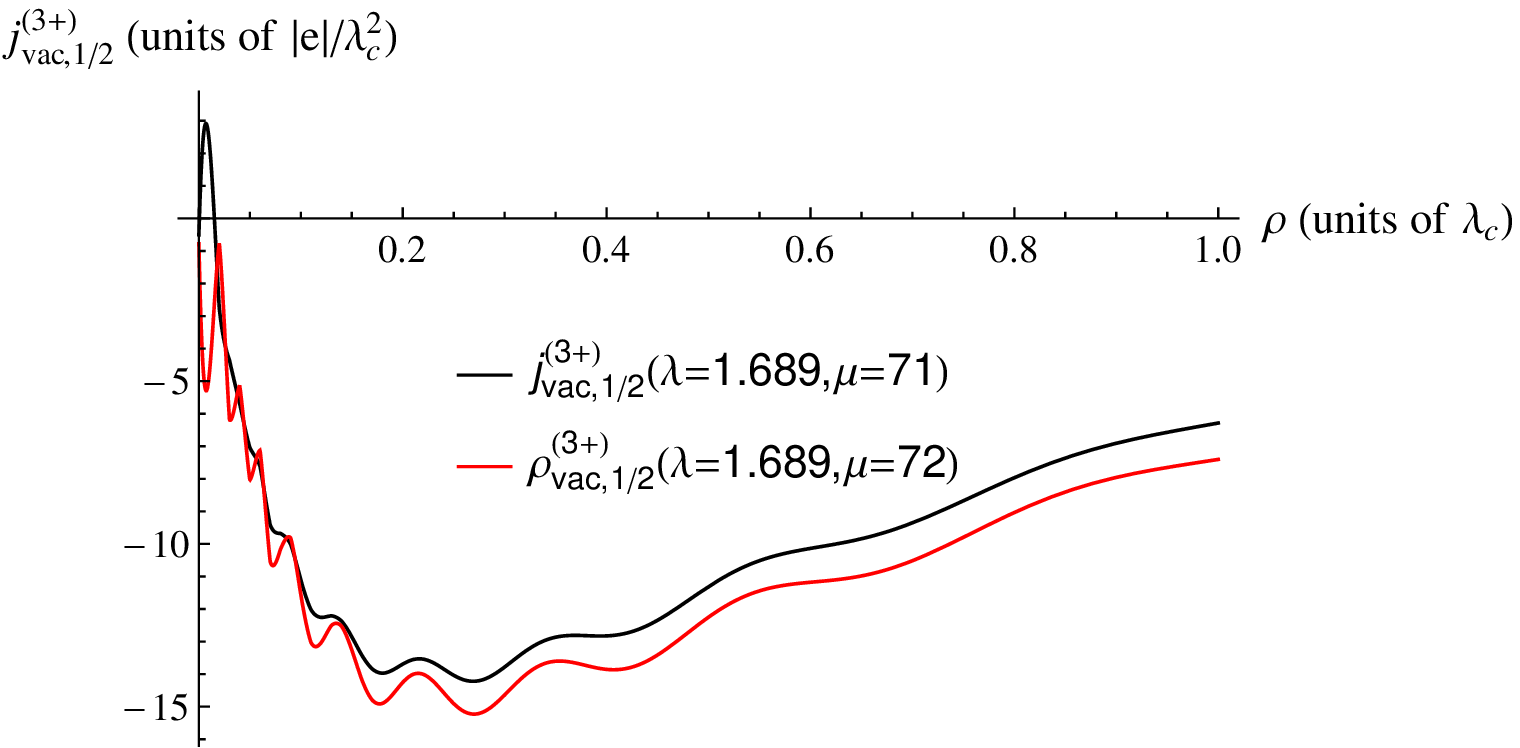}
}
\caption{The jumps in the induced charge and current densities due to discrete level diving into the lower continuum by passing $\m$ through certain critical value $\m_{cr}$ for $Z=10$, $\a=0.4$, $R_0=1/30$, $\l=1.689$ and  $71<\m_{cr}<72$.}
	\label{pic6}
\end{figure*}

Now let us turn to concrete calculations.
In the first step the area  of the seed current parameters $\l$ and $\m$, which allow  the self-consistent mode of induced current generation, should be determined. For these purposes a grid      in the parameter space was introduced, at each point of which the induced current and corresponding vector-potential were calculated. As a proximity measure between induced and seed currents and vector-potentials there was used the ratio of  induced $\m_{vac}$ to the seed $\m$ dipole magnetic moments, related to the corresponding currents via relation (\ref {2.3}). The results of this procedure for $Z=10$  are presented in Figs.\ref{pic2}a for $\a=0.4$ and \ref{pic2}b for $\a=0.8$.
\begin{figure*}[ht!]
\subfigure[]{
		\includegraphics[scale=0.55]{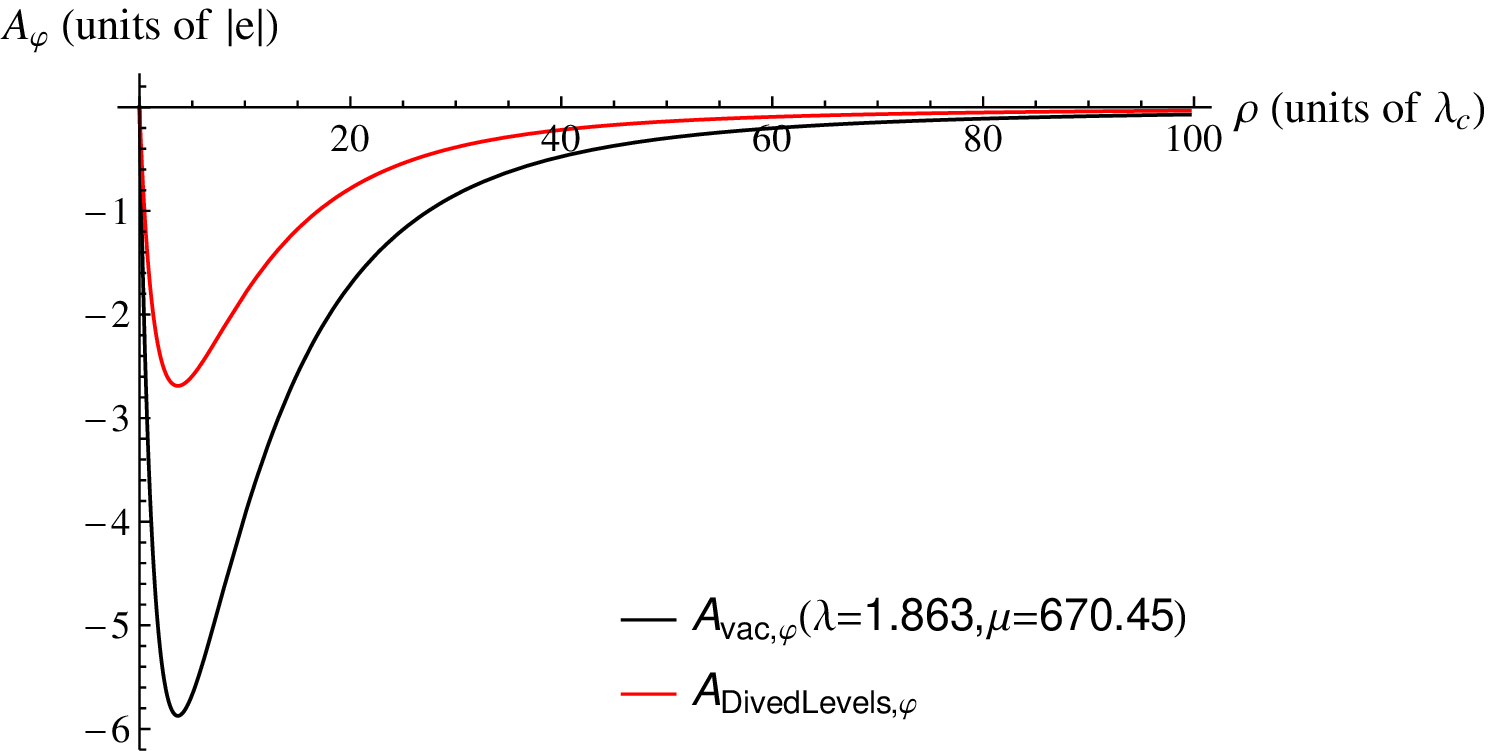}
}
\hfill
\subfigure[]{
		\includegraphics[scale=0.55]{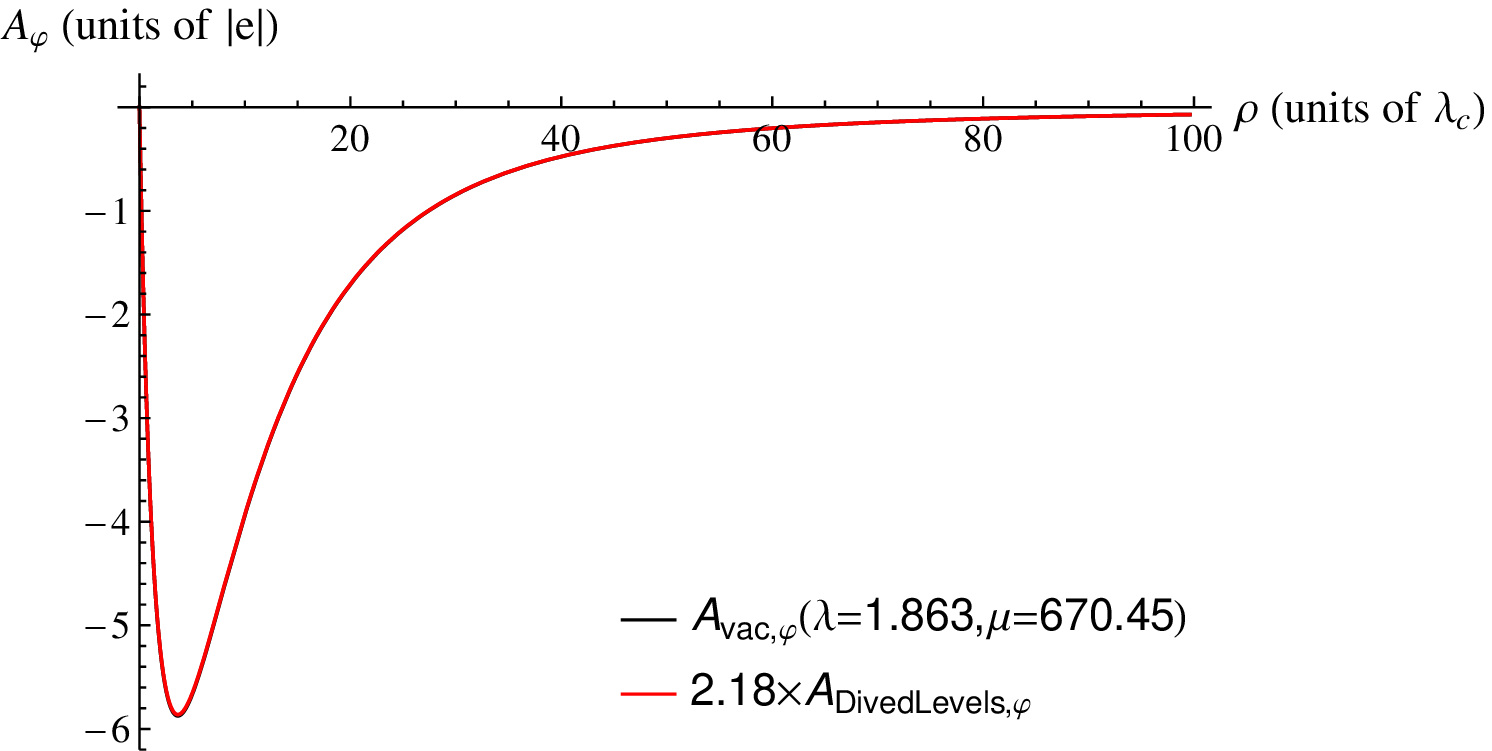}
}
\caption{Comparison between vector-potentials, created by  the total Dirac current generated by levels at the threshold of the lower continuum, with the exact self-consistent one for $Z=10$, $\a=0.8$, $R_0=1/175$, $\l=1.863$ and $\m=670.45$.}
	\label{pic5}
\end{figure*}
\begin{table*}[ht!]
\caption{Relationship between the seed, exact self-consistent and generated by the dived levels vector-potentials for  $Z=10$, $\a=0.8$, $R_0=1/175$, $\l=1.863$ and $\m=670.45$.}
\begin{center}
	\begin{tabular}{|c|c|c|c|c|c|c|c|c|c|c|}
		\hline $\r$ &1 & 2& 5& 10&15&20&30&40&50&100 \\[-0pt]
		\hline $A_\vf(\r)$ & -3.7409& -5.2661& -5.6696& -3.9364& -2.5669& -1.7100& -0.8438&-0.4753&-0.29762&-0.06906   \\
		\hline $A_{vac,\vf}^{ren}(\r)$ & -3.7423& -5.2679& -5.6712& -3.9375&-2.5678&-1.7105&-0.8442&-0.4755&-0.29769&-0.06908  \\
		\hline $2.1776\times$ & -3.7401& -5.2653& -5.6691& -3.9361&-2.5670&-1.7102&-0.8439&-0.4754&-0.29763&-0.06906  \\[-4pt]
		$A_{DivedLevels}(\r)$&&&&&&&&&&\\
		\hline
	\end{tabular}\label{t132}
\end{center}
\end{table*}

\begin{figure*}[ht!]
\subfigure[]{
		\includegraphics[scale=0.55]{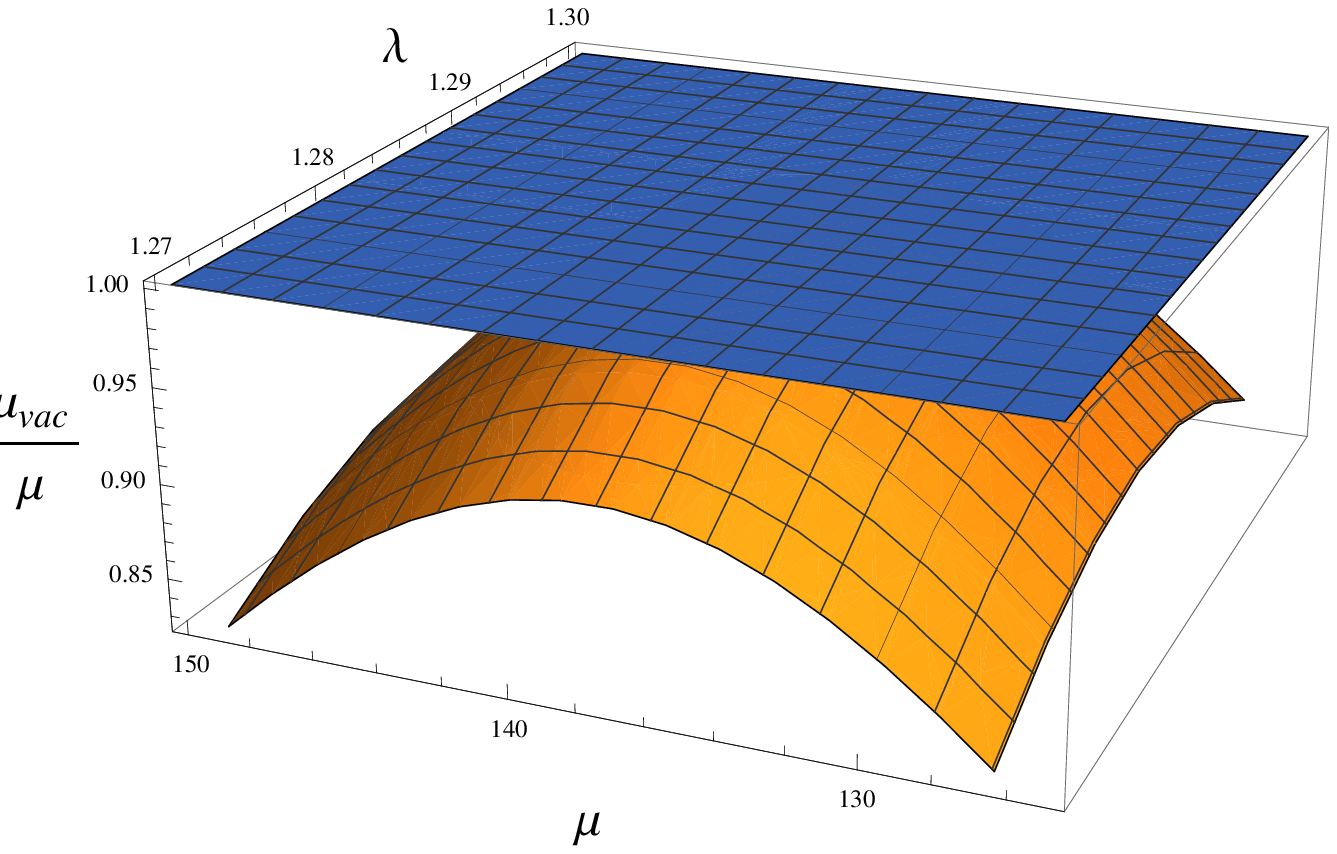}
}
\hfill
\subfigure[]{
		\includegraphics[scale=0.55]{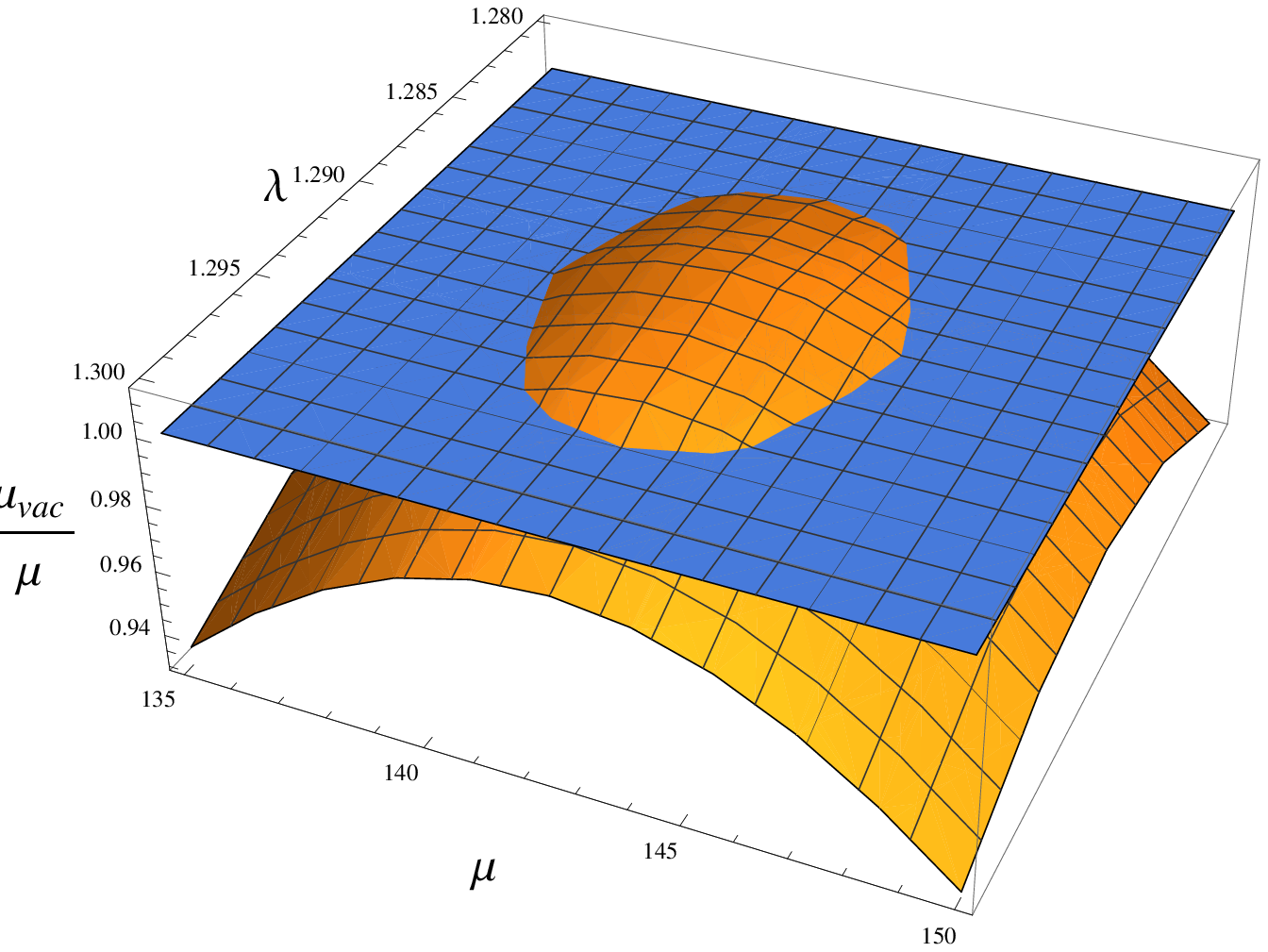}
}
\caption{The ratio $\m_{vac}/\m $ between the magnetic moment of exact induced current and the seed one for graphene on the SiC substrate with $\a=0.4$ and  $R_0=1/15$ for (a) $Z=4.625$ and (b) $Z=4.64$. The additional plane  corresponds to $\m_{vac}/\m=1$.}
	\label{pic7}
\end{figure*}

Figs.\ref{pic2} show that for both substrates there exist the whole sets of values for $\l$ and $\m$ in the form of closed loops in the parameter space, which provide $\m_{vac}=\m$, and hence,  the possibility of self-consistent generation of the induced current. Very instructive is also   the behavior of the induced current and corresponding  vector-potential with growing value of magnetic moment $\m$ of the seed current  for fixed  $\l$, which is also clearly seen in Figs.\ref{pic2}. First with growing $\m$ the ratio $\m_{vac}/\m$ increases monotonically until $\m$  reaches a  "saddle point"  $\m^{\ast}$, which depends on the current value of $\l$. After the parameter $\m $  has exceeded this point, with further growth of  $\m$ beyond $\m^{\ast}$ the ratio  $\m_{vac}/\m$ decreases.

Such behavior of the induced current with growing $\m$ reflects the same features in evolution of discrete levels. In Figs.\ref{pic3} the trajectories of the lowest discrete levels in the partial channel with $|m_j|=1/2$  are shown for $Z=10$, $\a=0.8$, $R_0=1/175$, $\l=1.863$ by varying  $\m$ in the vicinity of the "saddle point"  $\m^{\ast}$, which in this case lies in the interval $662 < \m^{\ast}< 663$. For  $\m <\m^{\ast}$ the increase of $\m$ leads to accelerated lowering of the levels towards the lower continuum, and hence, to growing number of discrete levels, dived into the latter, whereas for  $\m > \m^{\ast}$  --- vice versa. This is the reason for such peculiar behavior of the ratio $\m_{vac}/\m$ with growing $\m$. It would be also worth to note that for those values of $\l$ and $\m$, which provide the self-consistent mode of induced current generation, the total number of discrete levels, dived into the lower continuum, turns out to be substantially larger compared to the case without magnetic field. More explicitly this circumstance is shown in Figs.\ref{pic4}, where  there are presented the histograms of per partial channel distribution of the number of dived levels with (orange) and without (turquoise) magnetic field for substrates SiC and h-BN, respectively.

In Figs.\ref{pic6} the jump-like behavior of the induced  charge and current densities  by varying $\m$ in the vicinity of the critical value $\m_{cr}$, at which the level dives into the lower continuum, are presented. For more clarity there are shown only the $(3+)$-components of densities, since the diving effect is completely non-perturbative and therefore reveals primarily in these components. As expected, the value of the integral induced charge  $Q^{ren}_{vac}=\int d^2 \v{\r}\, \vr_{vac}^{ren}\(\v{\r}\)$ changes by $(-|e|)$, which can be easily verified via direct numerical calculation.

The assumption that the effect of spontaneous axial current generation is caused primarily by the dived levels, can be confirmed also in the next way. First for each discrete level from the set of dived into  the lower continuum, the corresponding Dirac current $\v{j}(\v{\r})=\p^{\dagger}(\v{\r})\,\v{\a}\,\p^{\dagger}(\v{\r})$ is evaluated, wherein the pertinent wavefunctions are taken at the lower threshold $\e=-1$. Thereafter one finds the sum of all such currents and evaluates the vector-potential, generated by them. In Fig.\ref{pic5}a the result of such procedure is presented for $\a=0.8$, $R_0=1/175$, $\l=1.863$ and $\m=670.45$. For comparison in  Fig.\ref{pic5}a the exact  induced vector-potential for the same  $\l$ and  $\m$ is also shown. It is easy to see  that the profile functions of these potentials are almost the same and differ only in magnitude. Namely, in Fig.\ref{pic5}b the exact induced vector-potential and the one, generated by the dived levels and multiplied by the factor $\simeq2.18$, are presented. There is no visible difference in them. The degree of coincidence between these two potentials  can be estimated from the Table~\ref{t132}, where the values of the corresponding functions  are given for a quite representative set of  $\r$'s. This means that the magnetic excitation of the electron-positron continuum, caused by the seed potential (\ref {2.5}), reproduces almost exactly the form of the latter.  Thus, due to correct choice of the initial approximation, already at first iteration  a  well-pronounced correspondence between the seed and induced vector-potentials is achieved. The successive iterations scheme in this case turns out to be rapidly converging, and so one can safely confine to a few first iterations.

Since the main contribution to the effect of spontaneous axial  current generation is caused primarily by the dived levels, there appears a natural question, how much dived levels are needed for its startup. Speaking otherwise, there  should exist certain  value of the impurity charge $Z^\ast$, which depends on the other system parameters and serves as a specific analogue of the Curie point in ferromagnetics. The latter implies that the self-consistency condition of axial current generation can be fulfilled   only when  $Z\geq Z^\ast$. In Figs.\ref{pic7} there are  presented the results of calculation the ratio $\m_{vac}/\m$  for graphene on the SiC substrate with  $\a=0.4$ and $R_0=1/15$ for $Z=4.625$ and $Z=4.64$, respectively.  Here the radius of the Coulomb source is taken equal to $R_0=2\,a$ for the most visual demonstration of the existence of a touch point, which denotes the startup of the self-consistent mode of the induced current generation.  For $Z=4.625$ there holds $\m_{vac}/\m < 1$ for any values of $\l$ and $\m$, while for $Z=4.64$ there appears already  a small set of  $\l$ and $\m$, which provide the existence of self-consistent mode. So for graphene on the  SiC substrate with $\a=0.4$ and $R_0=1/15$ one finds for the Curie point analogue the estimate $4.625<Z^\ast<4.64$, which is approximately twice the first critical charge $Z_{cr,1}\simeq 2.373$ for the purely Coulomb system with the same $\a$ and $R_0$ \cite{Voronina2019a}.

\begin{figure}
\center
\includegraphics[scale=0.6]{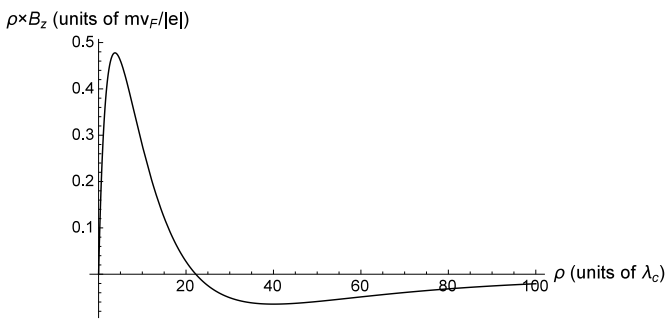}
\caption{\small The induced magnetic field $B_z$ spatial distribution for graphene on the SiC substrate doped by charged impurity with $R_0=1/15$ and $Z=5$, i.e. just beyond the Curie point $4.625<Z^\ast<4.64$. For more clarity we show the product $\r \times B_z$, otherwise the negative field amplitudes in the outer area are negligibly small compared to the (positive) inner ones.}
\label{pic8}
\end{figure}

Once the Curie point is reached, the spontaneous generation of the induced axial  current and associated magnetic dipole takes place. Fig.\ref{pic8} represents the induced magnetic field $B_z$ spatial distribution for graphene on the SiC substrate doped by charged impurity with $R_0=1/15$ and $Z=5$, i.e. just beyond the Curie point considered above. For such impurity parameters the polarization (vacuum) energy minimum is reached for $\lambda=1.347, \ \mu=165.34$ (the last  procedure is discussed in detail below in  Sect.IV). The  maximal (positive) amplitudes of the magnetic field are about $750$ Gauss and localize in a small vicinity of the circle with radius $\simeq 3.11$ nm.
The magnetic field  is positive inside the circle with radius $\simeq 18.58$ nm. The corresponding magnetic flux through this region equals to $\simeq 3.2144 \times 10^{-6}$ Gauss$\times$cm$^2$, which is equivalent to $\simeq 7.762$ units of the magnetic flux quantum unit $\mu_0=hc/|e|$. So for such impurity parameters the induced magnetic field is quite moderate and cannot significantly affect the main properties of the graphene plane, e.g., sample conductivity. Note also that the non-integer value of magnetic flux quanta through this region is a specific feature of the QED-vacuum polarization, since it proceeds without real charge carriers. The detailed explanation of this phenomenon is given in Refs.~\cite{Greiner1985a,Greiner2012,Plunien1986}. At the same time, the total magnetic flux through the whole graphene plane vanishes exactly due to the dipole-like structure of the induced magnetic field, that is clearly seen in Fig.\ref{pic8} and confirmed by direct calculation.

\section{Casimir (vacuum)  energy with magnetic polarization effects}

The starting expression for the vacuum energy is quite analogous to the induced charge and current densities  with the only principal difference that the energy should be further normalized on the free case~\cite{Sveshnikov2017, Greiner2012, Plunien1986}
\begin{multline}
\label{6.1}
\E_{vac}={1 \over 2} \(\sum\limits_{\e_n<\e_F} \e_n-\sum\limits_{\e_n \geqslant \e_F} \e_n
\)_A \ - \\  - \ {1 \over 2} \(\sum\limits_{\e_n < 0} \e_n-\sum\limits_{\e_n > 0} \e_n \)_0 \ ,
\end{multline}
where the label A denotes the case with external fields, while 0 stands for the free case.
\begin{figure*}[ht!]
\subfigure[]{
		\includegraphics[scale=0.48]{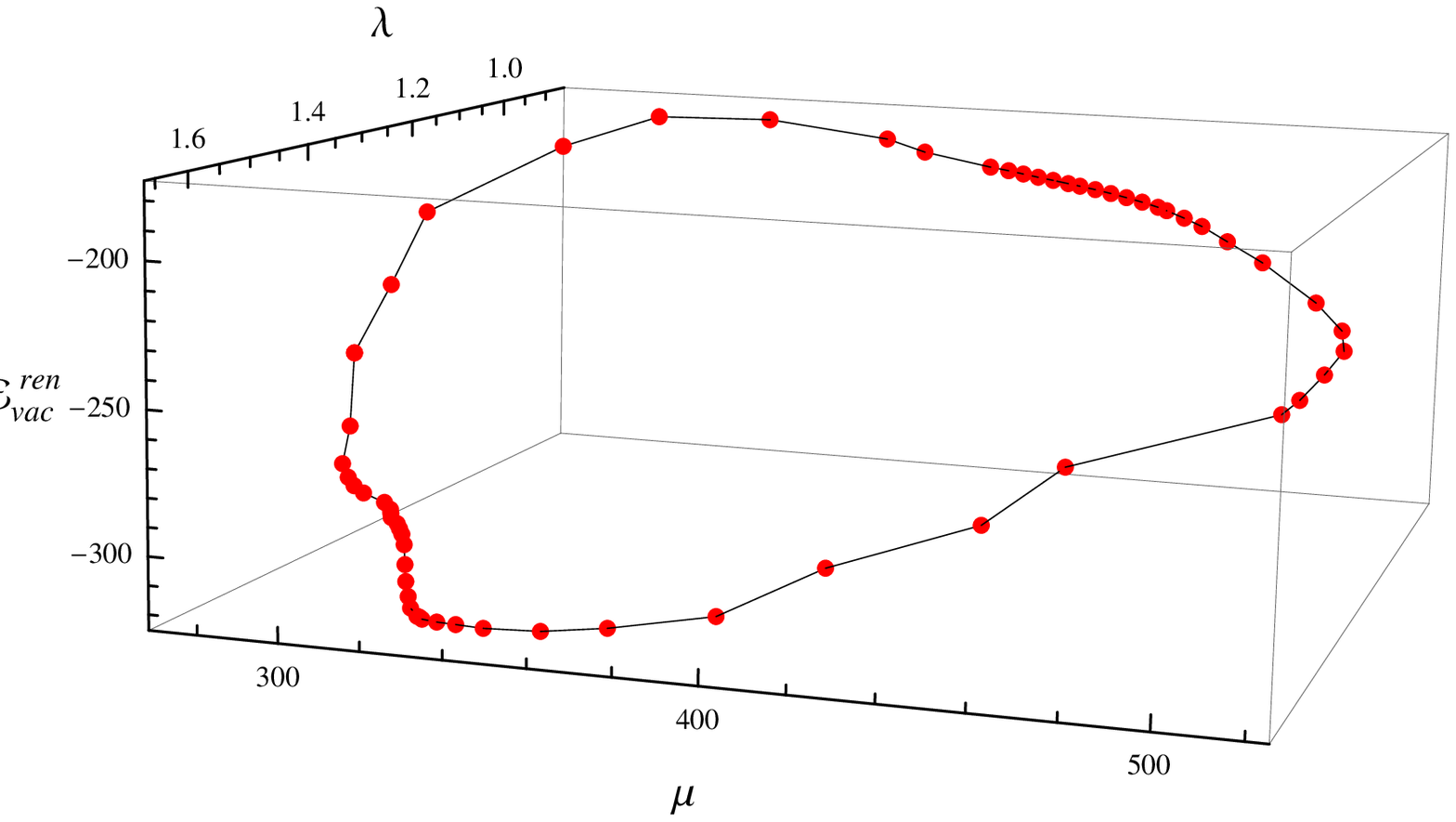}
}
\hfill
\subfigure[]{
		\includegraphics[scale=0.48]{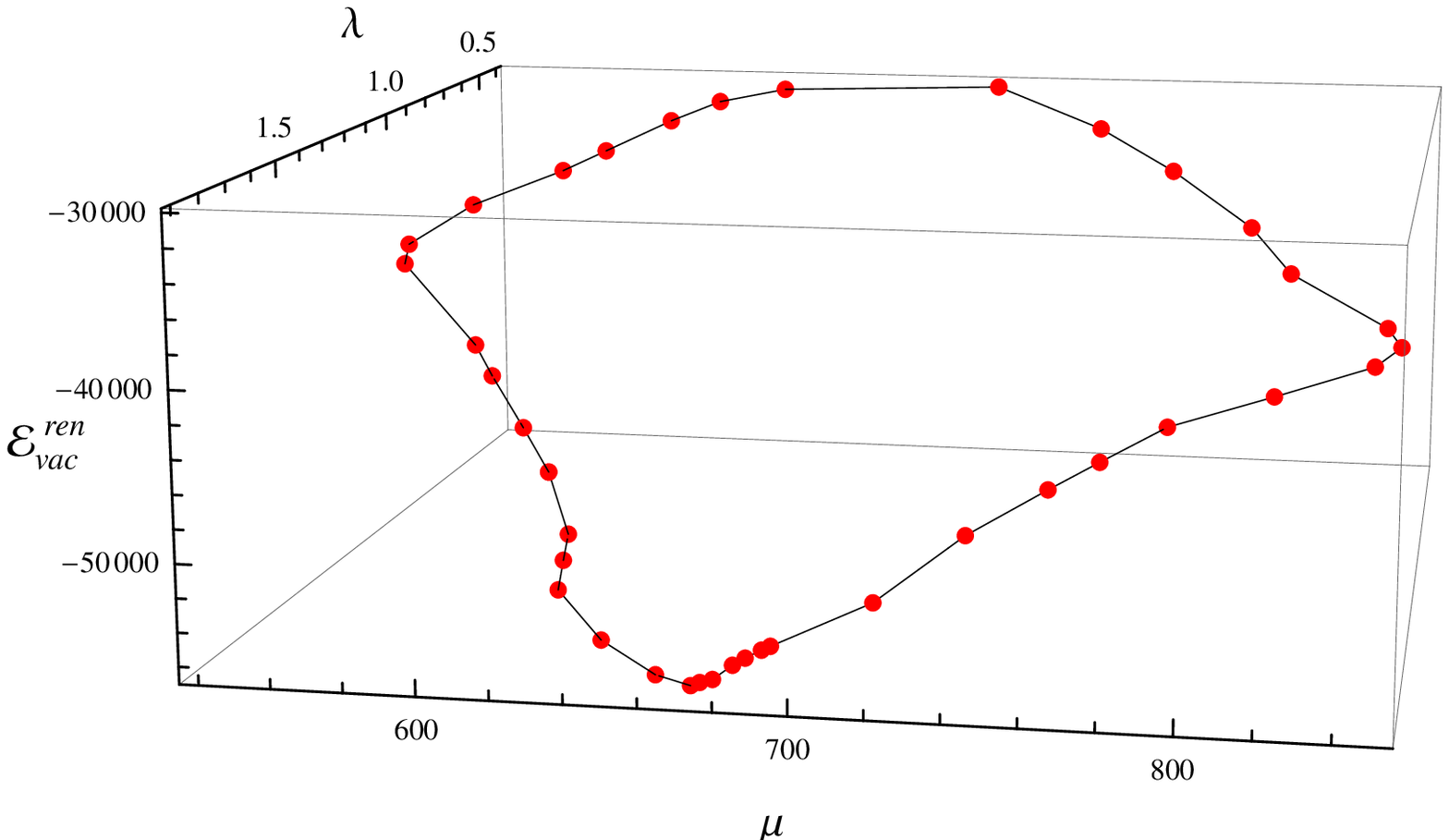}
}
\caption{The renormalized vacuum polarization energy for those sets of $\l$ and $\mu$, which provide the self-consistent mode of induced current generation for graphene:   (a) on the SiC substrate with $\a=0.4$ and $R_0=1/30$; (b) on the  h-BN substrate with $\a=0.8$ and $R_0=1/175$.}
	\label{pic9}
\end{figure*}

Since in the problem with magnetic field the analytic solutions of  DE are absent, an  alternative $\ln$[Wronskian] techniques, described in Refs.~\cite{Voronina2019c,Voronina2019d}, is applied. Within this approach the evaluation of the vacuum energy proceeds as follows
\beq
\label{6.2}
\E_{vac}=\sum_{m_j=1/2,\, 3/2,...}\E_{vac,|m_j|} \ ,
\eeq
\begin{widetext}
\beq
\label{6.3}
\E_{vac,|m_j|}=\sum_{s=ad,bc}\Bigg(\sum_{m_j=\pm |m_j|}{1 \over \pi}\int_0^\infty dy \, \mathrm{Re}\[\({y \over J^s_{m_j}(iy)}{dJ^s_{m_j}(iy) \over dy}\)_A-\({y \over J^s_{m_j}(iy)}{dJ^s_{m_j}(iy) \over dy}\)_0\] \ -  \ \sum_{-1\leq\e^s_{n,m_j}<0}\e^s_{n,m_j}\Bigg) \ ,
\eeq
\end{widetext}
where $J^s_{m_j}(iy)$ are the Wronskians (\ref{4.12}).

Since $\E_{vac}$ similar to $\vr_{vac}$ and $j_{vac}$ is represented by the partial expansion in $m_j$, there appears a problem of its convergence. In Refs.\cite{Davydov2018b, Voronina2019b, Sveshnikov2019b} there was shown that in absence of magnetic field this series in $m_j$ is linearly divergent. Further on account of model-independent considerations it was argued that  the elimination of this divergence should follow the same rules as in the QED-perturbation theory (PT) via regularization  the fermionic loop with two external lines.   Since in this case the vacuum energy to the first order of PT  consists of  two terms, which are quadratic either in $Z$ or in $\m$, its renormalization  proceeds now in the next way
\beq
\label{6.4a}
\E^{ren}_{vac}=\E_{vac,el}^{(1)} \   + \ \E_{vac,magn}^{(1)} \  +  \ \sum_{m_j=1/2,\, 3/2,...}\E^{ren}_{vac,|m_j|}  \ ,
\eeq
where
\begin{widetext}
\begin{multline}\label{6.4}
\E_{vac,|m_j|}^{ren}=\sum_{m_j=\pm |m_j|}\Bigg({1 \over \pi}\int_0^\infty dy \, \Bigg\{\mathrm{Re}\Bigg[\sum_{s=ad,bc}\({y \over J^s_{m_j}(iy)}{dJ^s_{m_j}(iy) \over dy}\Bigg|_A - {y \over J^s_{m_j}(iy)}{dJ^s_{m_j}(iy) \over dy}\Bigg|_0 \)\Bigg] \ - \ M_{B,|m_j|}(y)\Bigg\} \ - \\ - \ E_{B,|m_j|} \ -  \ \sum_{s=ad,bc}\sum_{-1\leq\e_{n,m_j}<0}\e^s_{n,m_j}\Bigg)  \ ,
\end{multline}
\end{widetext}
$\E_{vac,el}^{(1)}$ and $\E_{vac,magn}^{(1)}$ are the quadratic in $Z$ and $\mu$, correspondingly, perturbative   vacuum polarization energies
\beq\begin{gathered}
\label{3.11}
\E_{vac,el}^{(1)} = {1 \over 2}\int d^2\v{\r}^\prime\, \vr_{vac}^{(1)}(\v{\r}^\prime)\,A_0(\v{\r}^{\prime})  \ ,
\\
\E_{vac,magn}^{(1)}= -{1 \over 2} \int d^2\v{\r}^{\prime}\,\v{j}_{vac}^{(1)}(\v{\r}^\prime)\,\v{A}(\v{\r}^\prime) \ ,
\end{gathered}\eeq
while $E_{B,|m_j|}$ is the electric Born (quadratic in $Z$) component, which is found via first Born approximation for $\vr_{vac}(\r)$
\beq
\label{6.7}
E_{B,|m_j|}={|e| \over 2 \pi}\int_0^\infty d\r\, A_0(\r)\,\int_0^\infty dy\, \mathrm{Re}\[ \mathrm{Tr} G_{|m_j|}^{(1)}(\r;iy)\] \ .
\eeq
In contrast to $E_{B,|m_j|}$, the magnetic Born (quadratic in $\mu$) term
\begin{multline}\label{6.4a}
M_{B,|m_j|}(y)= \\ - \ \(|e|/2\)\,\int_0^\infty d\r\, A_\vf(\r)\, \mathrm{Re}\[\mathrm{Tr} \[\a_\vf G_{|m_j|}^{(1)}(\r;iy)\]\]
\end{multline}
cannot be taken out of the integration sign over $dy$ in the expression (\ref{6.4}), since the leading term of its asymptotics for $y\to\infty$ is $O(1/y)$, which cancels with the corresponding one of the Wronskians logarithmic derivatives, and so the leading term in the asymptotics of the integrand  in (\ref{6.4}) is  $O(1/y^2)$.

The results of Casimir energy evaluation for both substrates are presented in Figs.\ref{pic9} for $Z=10$ and those sets of $\l$ and $\m$, which form the closed loops shown in Figs.\ref{pic2} and so are able to provide the self-consistent mode of induced current generation. The main result is  that on those loops in the parameter space the vacuum energy reveals a well-pronounced minimum: for SiC substrate it is achieved for $\l=1.689$ and $\m=373.69$, while for h-BN substrate it occurs for   $\l=1.863$ and $\m=670.45$. The corresponding  values of vacuum energy turn out to be $\E_{vac}^{ren}=-674.167$ and $\E_{vac}^{ren}=-56518.3$, respectively, and lie substantially lower than in absence of magnetic field ($\E_{vac}^{ren}=-456.759$ and $\E_{vac}^{ren}=-25240.2$, see Ref.~\cite{Voronina2019b}). Therefore the spontaneous generation of ferromagnetic phase beyond the "Curie point" $Z \geq Z^{\ast}$  turns out to be energetically favorable compared to the purely electrostatic polarization in graphene on substrates SiC and h-BN.
 The latter circumstance confirms the possibility  of such purely non-perturbative magnetic effects in  planar electron-positron systems with strong coupling.\footnote{The recently discussed I- and Na-modified graphene grown on the Ir(111) surface, which reveals a very large unconventional gap that can be described in terms of a phenomenological massless Dirac model~\cite{Cappelluti2014}, lies beyond the scope of the present work.}

\section{Conclusion}

To conclude,  it is worth-while noticing that the   formation of the  ferromagnetic phase beyond the "Curie point" $Z \geq Z^{\ast}>Z_{cr,1}$ is nothing else but a nontrivial  example of spontaneous symmetry breaking. In the purely Coulomb case all the discrete levels are doubly degenerated with respect to the sign of $m_j$. Therefore upon diving of such doubly degenerated levels into the lower continuum the induced  current cannot appear, since the corresponding currents, created by states with opposite signs of $m_j$, compensate each other. At the same time, in presence of magnetic field this symmetry is broken. As it was shown in this work, for planar QED-systems similar to graphene  in presence of impurity with charge  $Z \geq Z^{\ast}$ there could arise an energetically more favorable  ferromagnetic  state, wherein the symmetry is spontaneously broken due to generation of a self-consistent induced axial current and corresponding magnetic dipole-like field, which splits the levels with opposite sign of $m_j$. The  value $Z^{\ast}$ of the Coulomb source, beyond which this effect could take place, significantly depends on the system parameters and has the meaning quite similar to  the Curie point in ferromagnetics.

The magnitudes of the dipole moment, induced beyond the "Curie point", turn out to be quite moderate and cannot significantly influence the main properties of the graphene plane, e.g., sample conductivity. At the energy minimum one has $\m_{vac}\simeq 373.69$ for $\a=0.4\, , R_0=1/30$ and $\m_{vac}\simeq 670.45$ for  $\a=0.8\, , R_0=1/175$, which in common units means $\m_{vac}\simeq 7.041 \times 10^5 \, \m_B\simeq 6.5 \times 10^{-15}$ erg/Gauss and $\m_{vac}\simeq 7.369 \times 10^6\, \m_B \simeq 6.8 \times 10^{-14}$ erg/Gauss.  The  specifics of such ferromagnetic phase is also that the spatial localization of the induced current and so of  magnetic dipole reproduces the corresponding one for the seed current (\ref {2.2}) and can be estimated as  $\sim \r\, \mathrm{e}^{-\l\sqrt{\r}}$. This means that  the induced ferromagnetic excitation should be much more distributed throughout the system volume compared to the purely exponential decay. In the example considered at the end of Sect.III the effective size of magnetic dipole is  dozens of nm (see Fig.\ref{pic8}). Similar spatial distribution is also the specific feature of the induced charge density.

The artificial creation of  charged impurities in graphene is a highly nontrivial task. The most part of such kind experiments deal with ionized adatoms with charge $+|e|$, which do not reach the  over-critical region. In experiments, described in Refs.~\cite{Wang2012},\cite{Wang2013}  and \cite{Wang2015}, there were used a trimer, a cluster of dimers,  and a cluster of adatoms correspondingly. In  Ref.~\cite{Wang2013} it was shown that achieving  critical $Z$ requires  the creation of clusters containing a large number of ions.
A quite interesting novel approach to simulation of a charged impurity in graphene is proposed in Ref.~\cite{Mao2016}. There was shown that a vacancy in graphene can possess a stable positive charge, which can be continuously increased by applying voltage pulses from the STM-needle. Such techniques allows to observe the evolution of the system during the transition from subcritical phase to the supercritical one.  Comparison of results with those of preceding studies on this subject \cite{Wang2012},\cite{Wang2013},\cite{Brar2011}, where the supercritical Coulomb source  was simulated either via adatoms or clusters of charged molecules, confirms that the structure created this way really shows up as a supercritical charged impurity.

The attention paid in the present paper to the  h-BN substrate is caused by the next reasons. First, the heterogeneity of local charge density in graphene on the h-BN substrate is 1-2 orders of magnitude less than in graphene on the standard substrate $\mathrm{SiO_2}$, which has been reliably confirmed in works \cite{Decker2011},\cite{Burson2013} by means of STM and kelvin probe microscopy. As a result, the graphene electronic devices on the h-BN substrate in a number of parameters significantly surpass those  on $\mathrm{SiO_2}$ \cite{Dean2010}, therefore the h-BN substrate is of a great practical interest. Apart this, the effective fine structure constant in graphene on the h-BN substrate is quite large $\a_g \simeq 0.8$. So the effects of super-criticality should be observed for quite moderate impurity charges $Z\sim 2$. As an additional option there was considered graphene on the   SiC substrate, since the latter has been already explored in earlier works on this subject \cite{Pereira2008}. Moreover, for some practical uses graphene on the   SiC substrate turns out to be more preferable compared to  $\mathrm{SiO_2}$ and h-BN \cite{Rengel2015}, since $\a_g$ for this substrate is twice less than for  h-BN and so the defects, which  serve as charged impurities,  have a smaller effect on sample conductivity.

\section{Acknowledgments}

The authors are very indebted to Dr. O.V.Pavlovsky and A.A.Krasnov from MSU Department of Physics and to A.S.Davydov from Kurchatov Center for interest and helpful discussions.  This work has been supported in part by the RF Ministry of Sc. $\&$ Ed.  Scientific Research Program, projects No. 01-2014-63889, A16-116021760047-5, and by RFBR grant No. 14-02-01261.

\bibliography{VP2DG}

\begin{thebibliography}{47}%
\makeatletter
\providecommand \@ifxundefined [1]{%
 \@ifx{#1\undefined}
}%
\providecommand \@ifnum [1]{%
 \ifnum #1\expandafter \@firstoftwo
 \else \expandafter \@secondoftwo
 \fi
}%
\providecommand \@ifx [1]{%
 \ifx #1\expandafter \@firstoftwo
 \else \expandafter \@secondoftwo
 \fi
}%
\providecommand \natexlab [1]{#1}%
\providecommand \enquote  [1]{``#1''}%
\providecommand \bibnamefont  [1]{#1}%
\providecommand \bibfnamefont [1]{#1}%
\providecommand \citenamefont [1]{#1}%
\providecommand \href@noop [0]{\@secondoftwo}%
\providecommand \href [0]{\begingroup \@sanitize@url \@href}%
\providecommand \@href[1]{\@@startlink{#1}\@@href}%
\providecommand \@@href[1]{\endgroup#1\@@endlink}%
\providecommand \@sanitize@url [0]{\catcode `\\12\catcode `\$12\catcode
  `\&12\catcode `\#12\catcode `\^12\catcode `\_12\catcode `\%12\relax}%
\providecommand \@@startlink[1]{}%
\providecommand \@@endlink[0]{}%
\providecommand \url  [0]{\begingroup\@sanitize@url \@url }%
\providecommand \@url [1]{\endgroup\@href {#1}{\urlprefix }}%
\providecommand \urlprefix  [0]{URL }%
\providecommand \Eprint [0]{\href }%
\providecommand \doibase [0]{http://dx.doi.org/}%
\providecommand \selectlanguage [0]{\@gobble}%
\providecommand \bibinfo  [0]{\@secondoftwo}%
\providecommand \bibfield  [0]{\@secondoftwo}%
\providecommand \translation [1]{[#1]}%
\providecommand \BibitemOpen [0]{}%
\providecommand \bibitemStop [0]{}%
\providecommand \bibitemNoStop [0]{.\EOS\space}%
\providecommand \EOS [0]{\spacefactor3000\relax}%
\providecommand \BibitemShut  [1]{\csname bibitem#1\endcsname}%
\let\auto@bib@innerbib\@empty
\bibitem [{\citenamefont {Katsnelson}(2006)}]{Katsnelson2006b}%
  \BibitemOpen
  \bibfield  {author} {\bibinfo {author} {\bibfnamefont {M.~I.}\ \bibnamefont
  {Katsnelson}},\ }\href {\doibase 10.1103/PhysRevB.74.201401} {\bibfield
  {journal} {\bibinfo  {journal} {Phys. Rev. B}\ }\textbf {\bibinfo {volume}
  {74}},\ \bibinfo {pages} {201401} (\bibinfo {year} {2006})}\BibitemShut
  {NoStop}%
\bibitem [{\citenamefont {Shytov}\ \emph {et~al.}(2007)\citenamefont {Shytov},
  \citenamefont {Katsnelson},\ and\ \citenamefont {Levitov}}]{Shytov2007}%
  \BibitemOpen
  \bibfield  {author} {\bibinfo {author} {\bibfnamefont {A.~V.}\ \bibnamefont
  {Shytov}}, \bibinfo {author} {\bibfnamefont {M.~I.}\ \bibnamefont
  {Katsnelson}}, \ and\ \bibinfo {author} {\bibfnamefont {L.~S.}\ \bibnamefont
  {Levitov}},\ }\href {\doibase 10.1103/PhysRevLett.99.236801} {\bibfield
  {journal} {\bibinfo  {journal} {Phys. Rev. Lett.}\ }\textbf {\bibinfo
  {volume} {99}},\ \bibinfo {pages} {236801} (\bibinfo {year}
  {2007})}\BibitemShut {NoStop}%
\bibitem [{\citenamefont {Kotov}\ \emph {et~al.}(2008)\citenamefont {Kotov},
  \citenamefont {Pereira},\ and\ \citenamefont {Uchoa}}]{Kotov2008}%
  \BibitemOpen
  \bibfield  {author} {\bibinfo {author} {\bibfnamefont {V.~N.}\ \bibnamefont
  {Kotov}}, \bibinfo {author} {\bibfnamefont {V.~M.}\ \bibnamefont {Pereira}},
  \ and\ \bibinfo {author} {\bibfnamefont {B.}~\bibnamefont {Uchoa}},\ }\href
  {\doibase 10.1103/PhysRevB.78.075433} {\bibfield  {journal} {\bibinfo
  {journal} {Phys. Rev. B}\ }\textbf {\bibinfo {volume} {78}},\ \bibinfo
  {pages} {075433} (\bibinfo {year} {2008})}\BibitemShut {NoStop}%
\bibitem [{\citenamefont {Pereira}\ \emph {et~al.}(2008)\citenamefont
  {Pereira}, \citenamefont {Kotov},\ and\ \citenamefont
  {Castro~Neto}}]{Pereira2008}%
  \BibitemOpen
  \bibfield  {author} {\bibinfo {author} {\bibfnamefont {V.~M.}\ \bibnamefont
  {Pereira}}, \bibinfo {author} {\bibfnamefont {V.~N.}\ \bibnamefont {Kotov}},
  \ and\ \bibinfo {author} {\bibfnamefont {A.~H.}\ \bibnamefont
  {Castro~Neto}},\ }\href {\doibase 10.1103/PhysRevB.78.085101} {\bibfield
  {journal} {\bibinfo  {journal} {Phys. Rev. B}\ }\textbf {\bibinfo {volume}
  {78}},\ \bibinfo {pages} {085101} (\bibinfo {year} {2008})}\BibitemShut
  {NoStop}%
\bibitem [{\citenamefont {Nishida}(2014)}]{Nishida2014}%
  \BibitemOpen
  \bibfield  {author} {\bibinfo {author} {\bibfnamefont {Y.}~\bibnamefont
  {Nishida}},\ }\href {\doibase 10.1103/PhysRevB.90.165414} {\bibfield
  {journal} {\bibinfo  {journal} {Phys. Rev. B}\ }\textbf {\bibinfo {volume}
  {90}},\ \bibinfo {pages} {165414} (\bibinfo {year} {2014})}\BibitemShut
  {NoStop}%
\bibitem [{\citenamefont {Bordag}\ \emph {et~al.}(2016)\citenamefont {Bordag},
  \citenamefont {Fialkovskiy},\ and\ \citenamefont {Vassilevich}}]{Bordag2016}%
  \BibitemOpen
  \bibfield  {author} {\bibinfo {author} {\bibfnamefont {M.}~\bibnamefont
  {Bordag}}, \bibinfo {author} {\bibfnamefont {I.}~\bibnamefont {Fialkovskiy}},
  \ and\ \bibinfo {author} {\bibfnamefont {D.}~\bibnamefont {Vassilevich}},\
  }\href {\doibase 10.1103/PhysRevB.93.075414} {\bibfield  {journal} {\bibinfo
  {journal} {Phys. Rev. B}\ }\textbf {\bibinfo {volume} {93}},\ \bibinfo
  {pages} {075414} (\bibinfo {year} {2016})}\BibitemShut {NoStop}%
\bibitem [{\citenamefont {Bordag}\ \emph {et~al.}(2017)\citenamefont {Bordag},
  \citenamefont {Fialkovskiy},\ and\ \citenamefont {Vassilevich}}]{Bordag2017}%
  \BibitemOpen
  \bibfield  {author} {\bibinfo {author} {\bibfnamefont {M.}~\bibnamefont
  {Bordag}}, \bibinfo {author} {\bibfnamefont {I.}~\bibnamefont {Fialkovskiy}},
  \ and\ \bibinfo {author} {\bibfnamefont {D.}~\bibnamefont {Vassilevich}},\
  }\href {\doibase 10.1103/PhysRevB.95.119905} {\bibfield  {journal} {\bibinfo
  {journal} {Phys. Rev. B}\ }\textbf {\bibinfo {volume} {95}},\ \bibinfo
  {pages} {119905} (\bibinfo {year} {2017})}\BibitemShut {NoStop}%
\bibitem [{\citenamefont {Khalilov}\ and\ \citenamefont
  {Mamsurov}(2017)}]{Khalilov2017}%
  \BibitemOpen
  \bibfield  {author} {\bibinfo {author} {\bibfnamefont {V.~R.}\ \bibnamefont
  {Khalilov}}\ and\ \bibinfo {author} {\bibfnamefont {I.~V.}\ \bibnamefont
  {Mamsurov}},\ }\href {\doibase 10.1016/j.physletb.2017.03.052} {\bibfield
  {journal} {\bibinfo  {journal} {Phys. Lett. B}\ }\textbf {\bibinfo {volume}
  {769}},\ \bibinfo {pages} {152} (\bibinfo {year} {2017})}\BibitemShut
  {NoStop}%
\bibitem [{\citenamefont {Voronina}\ \emph
  {et~al.}(2019{\natexlab{a}})\citenamefont {Voronina}, \citenamefont
  {Sveshnikov}, \citenamefont {Grashin},\ and\ \citenamefont
  {Davydov}}]{Voronina2019a}%
  \BibitemOpen
  \bibfield  {author} {\bibinfo {author} {\bibfnamefont {Y.}~\bibnamefont
  {Voronina}}, \bibinfo {author} {\bibfnamefont {K.}~\bibnamefont
  {Sveshnikov}}, \bibinfo {author} {\bibfnamefont {P.}~\bibnamefont {Grashin}},
  \ and\ \bibinfo {author} {\bibfnamefont {A.}~\bibnamefont {Davydov}},\ }\href
  {\doibase https://doi.org/10.1016/j.physe.2018.08.013} {\bibfield  {journal}
  {\bibinfo  {journal} {Physica E}\ }\textbf {\bibinfo {volume} {106}},\
  \bibinfo {pages} {298 } (\bibinfo {year} {2019}{\natexlab{a}})}\BibitemShut
  {NoStop}%
\bibitem [{\citenamefont {Voronina}\ \emph
  {et~al.}(2019{\natexlab{b}})\citenamefont {Voronina}, \citenamefont
  {Sveshnikov}, \citenamefont {Grashin},\ and\ \citenamefont
  {Davydov}}]{Voronina2019b}%
  \BibitemOpen
  \bibfield  {author} {\bibinfo {author} {\bibfnamefont {Y.}~\bibnamefont
  {Voronina}}, \bibinfo {author} {\bibfnamefont {K.}~\bibnamefont
  {Sveshnikov}}, \bibinfo {author} {\bibfnamefont {P.}~\bibnamefont {Grashin}},
  \ and\ \bibinfo {author} {\bibfnamefont {A.}~\bibnamefont {Davydov}},\ }\href
  {\doibase https://doi.org/10.1016/j.physe.2018.09.026} {\bibfield  {journal}
  {\bibinfo  {journal} {Physica E}\ }\textbf {\bibinfo {volume} {109}},\
  \bibinfo {pages} {209 } (\bibinfo {year} {2019}{\natexlab{b}})}\BibitemShut
  {NoStop}%
\bibitem [{\citenamefont {G{\'o}rnicki}(1990)}]{Gornicki1990}%
  \BibitemOpen
  \bibfield  {author} {\bibinfo {author} {\bibfnamefont {P.}~\bibnamefont
  {G{\'o}rnicki}},\ }\href {\doibase 10.1016/0003-4916(90)90226-E} {\bibfield
  {journal} {\bibinfo  {journal} {Ann. Phys.}\ }\textbf {\bibinfo {volume}
  {202}},\ \bibinfo {pages} {271} (\bibinfo {year} {1990})}\BibitemShut
  {NoStop}%
\bibitem [{\citenamefont {Milstein}\ and\ \citenamefont
  {Terekhov}(2011)}]{Milstein2011}%
  \BibitemOpen
  \bibfield  {author} {\bibinfo {author} {\bibfnamefont {A.~I.}\ \bibnamefont
  {Milstein}}\ and\ \bibinfo {author} {\bibfnamefont {I.~S.}\ \bibnamefont
  {Terekhov}},\ }\href {\doibase 10.1103/PhysRevB.83.075420} {\bibfield
  {journal} {\bibinfo  {journal} {Phys. Rev. B}\ }\textbf {\bibinfo {volume}
  {83}},\ \bibinfo {pages} {075420} (\bibinfo {year} {2011})}\BibitemShut
  {NoStop}%
\bibitem [{\citenamefont {Bordag}\ and\ \citenamefont
  {Kirsten}(1999)}]{Bordag1999}%
  \BibitemOpen
  \bibfield  {author} {\bibinfo {author} {\bibfnamefont {M.}~\bibnamefont
  {Bordag}}\ and\ \bibinfo {author} {\bibfnamefont {K.}~\bibnamefont
  {Kirsten}},\ }\href {\doibase 10.1103/PhysRevD.60.105019} {\bibfield
  {journal} {\bibinfo  {journal} {Phys. Rev. D}\ }\textbf {\bibinfo {volume}
  {60}},\ \bibinfo {pages} {105019} (\bibinfo {year} {1999})}\BibitemShut
  {NoStop}%
\bibitem [{\citenamefont {Jackiw}\ \emph {et~al.}(2009)\citenamefont {Jackiw},
  \citenamefont {Milstein}, \citenamefont {Pi},\ and\ \citenamefont
  {Terekhov}}]{Jackiw2009}%
  \BibitemOpen
  \bibfield  {author} {\bibinfo {author} {\bibfnamefont {R.}~\bibnamefont
  {Jackiw}}, \bibinfo {author} {\bibfnamefont {A.~I.}\ \bibnamefont
  {Milstein}}, \bibinfo {author} {\bibfnamefont {S.-Y.}\ \bibnamefont {Pi}}, \
  and\ \bibinfo {author} {\bibfnamefont {I.~S.}\ \bibnamefont {Terekhov}},\
  }\href {\doibase 10.1103/PhysRevB.80.033413} {\bibfield  {journal} {\bibinfo
  {journal} {Phys. Rev. B}\ }\textbf {\bibinfo {volume} {80}},\ \bibinfo
  {pages} {033413} (\bibinfo {year} {2009})}\BibitemShut {NoStop}%
\bibitem [{\citenamefont {Khalilov}\ and\ \citenamefont
  {Lee}(2012)}]{Khalilov2012}%
  \BibitemOpen
  \bibfield  {author} {\bibinfo {author} {\bibfnamefont {V.~R.}\ \bibnamefont
  {Khalilov}}\ and\ \bibinfo {author} {\bibfnamefont {K.~E.}\ \bibnamefont
  {Lee}},\ }\href {\doibase 10.1142/S0217751X12501692} {\bibfield  {journal}
  {\bibinfo  {journal} {Int. J. Mod. Phys. A}\ }\textbf {\bibinfo {volume}
  {27}},\ \bibinfo {pages} {1250169} (\bibinfo {year} {2012})}\BibitemShut
  {NoStop}%
\bibitem [{\citenamefont {Khalilov}(2013)}]{Khalilov2013}%
  \BibitemOpen
  \bibfield  {author} {\bibinfo {author} {\bibfnamefont {V.}~\bibnamefont
  {Khalilov}},\ }\href {\doibase
  http://dx.doi.org/10.1140/epjc/s10052-013-2548-x} {\bibfield  {journal}
  {\bibinfo  {journal} {Eur. Phys. J. C}\ }\textbf {\bibinfo {volume} {73}},\
  \bibinfo {pages} {2548} (\bibinfo {year} {2013})}\BibitemShut {NoStop}%
\bibitem [{\citenamefont {Khalilov}(2014)}]{Khalilov2014}%
  \BibitemOpen
  \bibfield  {author} {\bibinfo {author} {\bibfnamefont {V.}~\bibnamefont
  {Khalilov}},\ }\href {\doibase
  http://dx.doi.org/10.1140/epjc/s10052-014-3061-6} {\bibfield  {journal}
  {\bibinfo  {journal} {Eur. Phys. J. C}\ }\textbf {\bibinfo {volume} {74}},\
  \bibinfo {pages} {3061} (\bibinfo {year} {2014})}\BibitemShut {NoStop}%
\bibitem [{\citenamefont {Wichmann}\ and\ \citenamefont
  {Kroll}(1956)}]{Wichmann1956}%
  \BibitemOpen
  \bibfield  {author} {\bibinfo {author} {\bibfnamefont {E.~H.}\ \bibnamefont
  {Wichmann}}\ and\ \bibinfo {author} {\bibfnamefont {N.~M.}\ \bibnamefont
  {Kroll}},\ }\href {\doibase 10.1103/PhysRev.101.843} {\bibfield  {journal}
  {\bibinfo  {journal} {Phys. Rev.}\ }\textbf {\bibinfo {volume} {101}},\
  \bibinfo {pages} {843} (\bibinfo {year} {1956})}\BibitemShut {NoStop}%
\bibitem [{\citenamefont {Nishida}(2016)}]{Nishida2016}%
  \BibitemOpen
  \bibfield  {author} {\bibinfo {author} {\bibfnamefont {Y.}~\bibnamefont
  {Nishida}},\ }\href {\doibase 10.1103/PhysRevB.94.085430} {\bibfield
  {journal} {\bibinfo  {journal} {Phys. Rev. B}\ }\textbf {\bibinfo {volume}
  {94}},\ \bibinfo {pages} {085430} (\bibinfo {year} {2016})}\BibitemShut
  {NoStop}%
\bibitem [{\citenamefont {Hassanabadi}\ \emph {et~al.}(2013)\citenamefont
  {Hassanabadi}, \citenamefont {Maghsoodi},\ and\ \citenamefont
  {Zarrinkamar}}]{Hassanabadi2013}%
  \BibitemOpen
  \bibfield  {author} {\bibinfo {author} {\bibfnamefont {H.}~\bibnamefont
  {Hassanabadi}}, \bibinfo {author} {\bibfnamefont {E.}~\bibnamefont
  {Maghsoodi}}, \ and\ \bibinfo {author} {\bibfnamefont {S.}~\bibnamefont
  {Zarrinkamar}},\ }\href {\doibase 10.1002/andp.201300102} {\bibfield
  {journal} {\bibinfo  {journal} {Ann. der Physik}\ }\textbf {\bibinfo {volume}
  {525}},\ \bibinfo {pages} {387} (\bibinfo {year} {2013})}\BibitemShut
  {NoStop}%
\bibitem [{\citenamefont {Davydov}\ \emph {et~al.}(2017)\citenamefont
  {Davydov}, \citenamefont {Sveshnikov},\ and\ \citenamefont
  {Voronina}}]{Davydov2017}%
  \BibitemOpen
  \bibfield  {author} {\bibinfo {author} {\bibfnamefont {A.}~\bibnamefont
  {Davydov}}, \bibinfo {author} {\bibfnamefont {K.}~\bibnamefont {Sveshnikov}},
  \ and\ \bibinfo {author} {\bibfnamefont {Y.}~\bibnamefont {Voronina}},\
  }\href {\doibase 10.1142/S0217751X17500543} {\bibfield  {journal} {\bibinfo
  {journal} {Int. J. Mod. Phys. A}\ }\textbf {\bibinfo {volume} {32}},\
  \bibinfo {pages} {1750054} (\bibinfo {year} {2017})}\BibitemShut {NoStop}%
\bibitem [{\citenamefont {Voronina}\ \emph
  {et~al.}(2017{\natexlab{a}})\citenamefont {Voronina}, \citenamefont
  {Davydov},\ and\ \citenamefont {Sveshnikov}}]{Voronina2017}%
  \BibitemOpen
  \bibfield  {author} {\bibinfo {author} {\bibfnamefont {Y.}~\bibnamefont
  {Voronina}}, \bibinfo {author} {\bibfnamefont {A.}~\bibnamefont {Davydov}}, \
  and\ \bibinfo {author} {\bibfnamefont {K.}~\bibnamefont {Sveshnikov}},\
  }\href {\doibase 10.1134/S1547477117050144} {\bibfield  {journal} {\bibinfo
  {journal} {Phys. Part. Nucl. Lett.}\ }\textbf {\bibinfo {volume} {14}},\
  \bibinfo {pages} {698 } (\bibinfo {year} {2017}{\natexlab{a}})}\BibitemShut
  {NoStop}%
\bibitem [{\citenamefont {Voronina}\ \emph
  {et~al.}(2017{\natexlab{b}})\citenamefont {Voronina}, \citenamefont
  {Davydov},\ and\ \citenamefont {Sveshnikov}}]{Sveshnikov2017}%
  \BibitemOpen
  \bibfield  {author} {\bibinfo {author} {\bibfnamefont {Y.}~\bibnamefont
  {Voronina}}, \bibinfo {author} {\bibfnamefont {A.}~\bibnamefont {Davydov}}, \
  and\ \bibinfo {author} {\bibfnamefont {K.}~\bibnamefont {Sveshnikov}},\
  }\href {\doibase 10.1134/S004057791711006X} {\bibfield  {journal} {\bibinfo
  {journal} {Theor. Math. Phys.}\ }\textbf {\bibinfo {volume} {193}},\ \bibinfo
  {pages} {1647} (\bibinfo {year} {2017}{\natexlab{b}})}\BibitemShut {NoStop}%
\bibitem [{\citenamefont {Davydov}\ \emph
  {et~al.}(2018{\natexlab{a}})\citenamefont {Davydov}, \citenamefont
  {Sveshnikov},\ and\ \citenamefont {Voronina}}]{Davydov2018b}%
  \BibitemOpen
  \bibfield  {author} {\bibinfo {author} {\bibfnamefont {A.}~\bibnamefont
  {Davydov}}, \bibinfo {author} {\bibfnamefont {K.}~\bibnamefont {Sveshnikov}},
  \ and\ \bibinfo {author} {\bibfnamefont {Y.}~\bibnamefont {Voronina}},\
  }\href {\doibase 10.1142/S0217751X18500057} {\bibfield  {journal} {\bibinfo
  {journal} {Int. J. Mod. Phys. A}\ }\textbf {\bibinfo {volume} {33}},\
  \bibinfo {pages} {1850005} (\bibinfo {year}
  {2018}{\natexlab{a}})}\BibitemShut {NoStop}%
\bibitem [{\citenamefont {Sveshnikov}\ \emph
  {et~al.}(2019{\natexlab{a}})\citenamefont {Sveshnikov}, \citenamefont
  {Voronina}, \citenamefont {Davydov},\ and\ \citenamefont
  {Grashin}}]{Sveshnikov2019b}%
  \BibitemOpen
  \bibfield  {author} {\bibinfo {author} {\bibfnamefont {K.}~\bibnamefont
  {Sveshnikov}}, \bibinfo {author} {\bibfnamefont {Y.}~\bibnamefont
  {Voronina}}, \bibinfo {author} {\bibfnamefont {A.}~\bibnamefont {Davydov}}, \
  and\ \bibinfo {author} {\bibfnamefont {P.}~\bibnamefont {Grashin}},\ }\href
  {\doibase doi.org/10.1134/S0040577919040056} {\bibfield  {journal} {\bibinfo
  {journal} {Theor. Math. Phys.}\ }\textbf {\bibinfo {volume} {199}},\ \bibinfo
  {pages} {533} (\bibinfo {year} {2019}{\natexlab{a}})}\BibitemShut {NoStop}%
\bibitem [{\citenamefont {Voronina}\ \emph
  {et~al.}(2019{\natexlab{c}})\citenamefont {Voronina}, \citenamefont
  {Komissarov},\ and\ \citenamefont {Sveshnikov}}]{Voronina2019c}%
  \BibitemOpen
  \bibfield  {author} {\bibinfo {author} {\bibfnamefont {Y.}~\bibnamefont
  {Voronina}}, \bibinfo {author} {\bibfnamefont {I.}~\bibnamefont
  {Komissarov}}, \ and\ \bibinfo {author} {\bibfnamefont {K.}~\bibnamefont
  {Sveshnikov}},\ }\href {\doibase https://doi.org/10.1016/j.aop.2019.02.014}
  {\bibfield  {journal} {\bibinfo  {journal} {Ann. Phys.}\ }\textbf {\bibinfo
  {volume} {404}},\ \bibinfo {pages} {132 } (\bibinfo {year}
  {2019}{\natexlab{c}})}\BibitemShut {NoStop}%
\bibitem [{\citenamefont {Voronina}\ \emph
  {et~al.}(2019{\natexlab{d}})\citenamefont {Voronina}, \citenamefont
  {Komissarov},\ and\ \citenamefont {Sveshnikov}}]{Voronina2019d}%
  \BibitemOpen
  \bibfield  {author} {\bibinfo {author} {\bibfnamefont {Y.}~\bibnamefont
  {Voronina}}, \bibinfo {author} {\bibfnamefont {I.}~\bibnamefont
  {Komissarov}}, \ and\ \bibinfo {author} {\bibfnamefont {K.}~\bibnamefont
  {Sveshnikov}},\ }\href {\doibase 10.1103/PhysRevA.99.062504} {\bibfield
  {journal} {\bibinfo  {journal} {Phys. Rev. A}\ }\textbf {\bibinfo {volume}
  {99}},\ \bibinfo {pages} {062504} (\bibinfo {year}
  {2019}{\natexlab{d}})}\BibitemShut {NoStop}%
\bibitem [{\citenamefont {Goerbig}(2011)}]{Goerbig2011}%
  \BibitemOpen
  \bibfield  {author} {\bibinfo {author} {\bibfnamefont {M.~O.}\ \bibnamefont
  {Goerbig}},\ }\href {\doibase 10.1103/RevModPhys.83.1193} {\bibfield
  {journal} {\bibinfo  {journal} {Rev. Mod. Phys.}\ }\textbf {\bibinfo {volume}
  {83}},\ \bibinfo {pages} {1193} (\bibinfo {year} {2011})}\BibitemShut
  {NoStop}%
\bibitem [{\citenamefont {Wallbank}\ \emph {et~al.}(2015)\citenamefont
  {Wallbank}, \citenamefont {Mucha-Kruczyński}, \citenamefont {Chen},\ and\
  \citenamefont {Fal'ko}}]{Wallbank2015}%
  \BibitemOpen
  \bibfield  {author} {\bibinfo {author} {\bibfnamefont {J.}~\bibnamefont
  {Wallbank}}, \bibinfo {author} {\bibfnamefont {M.}~\bibnamefont
  {Mucha-Kruczyński}}, \bibinfo {author} {\bibfnamefont {X.}~\bibnamefont
  {Chen}}, \ and\ \bibinfo {author} {\bibfnamefont {M.}~\bibnamefont
  {Fal'ko}},\ }\href {\doibase 10.1002/andp.201400204} {\bibfield  {journal}
  {\bibinfo  {journal} {Ann. der Physik}\ }\textbf {\bibinfo {volume} {527}},\
  \bibinfo {pages} {359} (\bibinfo {year} {2015})}\BibitemShut {NoStop}%
\bibitem [{\citenamefont {Sadeghi}\ \emph {et~al.}(2016)\citenamefont
  {Sadeghi}, \citenamefont {Sangtarash},\ and\ \citenamefont
  {Lambert}}]{Sadeghi2015}%
  \BibitemOpen
  \bibfield  {author} {\bibinfo {author} {\bibfnamefont {H.}~\bibnamefont
  {Sadeghi}}, \bibinfo {author} {\bibfnamefont {S.}~\bibnamefont {Sangtarash}},
  \ and\ \bibinfo {author} {\bibfnamefont {C.}~\bibnamefont {Lambert}},\ }\href
  {\doibase 10.1016/j.physe.2015.09.005} {\bibfield  {journal} {\bibinfo
  {journal} {Physica E}\ }\textbf {\bibinfo {volume} {82}},\ \bibinfo {pages}
  {12 } (\bibinfo {year} {2016})}\BibitemShut {NoStop}%
\bibitem [{\citenamefont {Greiner}\ \emph {et~al.}(1985)\citenamefont
  {Greiner}, \citenamefont {M\"uller},\ and\ \citenamefont
  {Rafelski}}]{Greiner1985a}%
  \BibitemOpen
  \bibfield  {author} {\bibinfo {author} {\bibfnamefont {W.}~\bibnamefont
  {Greiner}}, \bibinfo {author} {\bibfnamefont {B.}~\bibnamefont {M\"uller}}, \
  and\ \bibinfo {author} {\bibfnamefont {J.}~\bibnamefont {Rafelski}},\ }\href
  {http://link.springer.com/book/10.1007/978-3-642-82272-8} {\emph {\bibinfo
  {title} {Quantum Electrodynamics of Strong Fields}}},\ \bibinfo {edition}
  {2nd}\ ed.\ (\bibinfo  {publisher} {Springer},\ \bibinfo {address} {Berlin},\
  \bibinfo {year} {1985})\BibitemShut {NoStop}%
\bibitem [{\citenamefont {Greiner}\ and\ \citenamefont
  {Reinhardt}(2009)}]{Greiner2012}%
  \BibitemOpen
  \bibfield  {author} {\bibinfo {author} {\bibfnamefont {W.}~\bibnamefont
  {Greiner}}\ and\ \bibinfo {author} {\bibfnamefont {J.}~\bibnamefont
  {Reinhardt}},\ }\href {\doibase /10.1007/978-3-540-87561-1} {\emph {\bibinfo
  {title} {Quantum Electrodynamics}}},\ \bibinfo {edition} {4th}\ ed.\
  (\bibinfo  {publisher} {Springer-Verlag Berlin Heidelberg},\ \bibinfo {year}
  {2009})\BibitemShut {NoStop}%
\bibitem [{\citenamefont {Gyulassy}(1975)}]{Gyulassy1975}%
  \BibitemOpen
  \bibfield  {author} {\bibinfo {author} {\bibfnamefont {M.}~\bibnamefont
  {Gyulassy}},\ }\href {\doibase 10.1016/0375-9474(75)90554-0} {\bibfield
  {journal} {\bibinfo  {journal} {Nucl. Phys. A}\ }\textbf {\bibinfo {volume}
  {244}},\ \bibinfo {pages} {497 } (\bibinfo {year} {1975})}\BibitemShut
  {NoStop}%
\bibitem [{\citenamefont {Davydov}\ \emph
  {et~al.}(2018{\natexlab{b}})\citenamefont {Davydov}, \citenamefont
  {Sveshnikov},\ and\ \citenamefont {Voronina}}]{Davydov2018a}%
  \BibitemOpen
  \bibfield  {author} {\bibinfo {author} {\bibfnamefont {A.}~\bibnamefont
  {Davydov}}, \bibinfo {author} {\bibfnamefont {K.}~\bibnamefont {Sveshnikov}},
  \ and\ \bibinfo {author} {\bibfnamefont {Y.}~\bibnamefont {Voronina}},\
  }\href {\doibase 10.1142/S0217751X18500045} {\bibfield  {journal} {\bibinfo
  {journal} {Int. J. Mod. Phys. A}\ }\textbf {\bibinfo {volume} {33}},\
  \bibinfo {pages} {1850004} (\bibinfo {year}
  {2018}{\natexlab{b}})}\BibitemShut {NoStop}%
\bibitem [{\citenamefont {Sveshnikov}\ \emph
  {et~al.}(2019{\natexlab{b}})\citenamefont {Sveshnikov}, \citenamefont
  {Voronina}, \citenamefont {Davydov},\ and\ \citenamefont
  {Grashin}}]{Sveshnikov2019a}%
  \BibitemOpen
  \bibfield  {author} {\bibinfo {author} {\bibfnamefont {K.}~\bibnamefont
  {Sveshnikov}}, \bibinfo {author} {\bibfnamefont {Y.}~\bibnamefont
  {Voronina}}, \bibinfo {author} {\bibfnamefont {A.}~\bibnamefont {Davydov}}, \
  and\ \bibinfo {author} {\bibfnamefont {P.}~\bibnamefont {Grashin}},\ }\href
  {\doibase doi.org/10.1134/S0040577919030024} {\bibfield  {journal} {\bibinfo
  {journal} {Theor. Math. Phys.}\ }\textbf {\bibinfo {volume} {198}},\ \bibinfo
  {pages} {331} (\bibinfo {year} {2019}{\natexlab{b}})}\BibitemShut {NoStop}%
\bibitem [{\citenamefont {Fano}(1961)}]{Fano1961}%
  \BibitemOpen
  \bibfield  {author} {\bibinfo {author} {\bibfnamefont {U.}~\bibnamefont
  {Fano}},\ }\href {\doibase 10.1103/PhysRev.124.1866} {\bibfield  {journal}
  {\bibinfo  {journal} {Phys. Rev.}\ }\textbf {\bibinfo {volume} {124}},\
  \bibinfo {pages} {1866} (\bibinfo {year} {1961})}\BibitemShut {NoStop}%
\bibitem [{\citenamefont {Plunien}\ \emph {et~al.}(1986)\citenamefont
  {Plunien}, \citenamefont {M\"uller},\ and\ \citenamefont
  {Greiner}}]{Plunien1986}%
  \BibitemOpen
  \bibfield  {author} {\bibinfo {author} {\bibfnamefont {G.}~\bibnamefont
  {Plunien}}, \bibinfo {author} {\bibfnamefont {B.}~\bibnamefont {M\"uller}}, \
  and\ \bibinfo {author} {\bibfnamefont {W.}~\bibnamefont {Greiner}},\ }\href
  {\doibase 10.1016/0370-1573(86)90020-7} {\bibfield  {journal} {\bibinfo
  {journal} {Phys. Rep.}\ }\textbf {\bibinfo {volume} {134}},\ \bibinfo {pages}
  {87 } (\bibinfo {year} {1986})}\BibitemShut {NoStop}%
\bibitem [{\citenamefont {Cappelluti}\ \emph {et~al.}(2014)\citenamefont
  {Cappelluti}, \citenamefont {Benfatto}, \citenamefont {Papagno},
  \citenamefont {Pacilè}, \citenamefont {Sheverdyaeva},\ and\ \citenamefont
  {Moras}}]{Cappelluti2014}%
  \BibitemOpen
  \bibfield  {author} {\bibinfo {author} {\bibfnamefont {E.}~\bibnamefont
  {Cappelluti}}, \bibinfo {author} {\bibfnamefont {L.}~\bibnamefont
  {Benfatto}}, \bibinfo {author} {\bibfnamefont {M.}~\bibnamefont {Papagno}},
  \bibinfo {author} {\bibfnamefont {D.}~\bibnamefont {Pacilè}}, \bibinfo
  {author} {\bibfnamefont {P.}~\bibnamefont {Sheverdyaeva}}, \ and\ \bibinfo
  {author} {\bibfnamefont {P.}~\bibnamefont {Moras}},\ }\href {\doibase
  10.1002/andp.201400123} {\bibfield  {journal} {\bibinfo  {journal} {Ann. der
  Physik}\ }\textbf {\bibinfo {volume} {526}},\ \bibinfo {pages} {387}
  (\bibinfo {year} {2014})}\BibitemShut {NoStop}%
\bibitem [{\citenamefont {Wang}\ \emph {et~al.}(2012)\citenamefont {Wang},
  \citenamefont {Brar}, \citenamefont {Shytov}, \citenamefont {Wu},
  \citenamefont {R.}, \citenamefont {Tsai}, \citenamefont {Zettl},
  \citenamefont {Levitov},\ and\ \citenamefont {Crommie}}]{Wang2012}%
  \BibitemOpen
  \bibfield  {author} {\bibinfo {author} {\bibfnamefont {Y.}~\bibnamefont
  {Wang}}, \bibinfo {author} {\bibfnamefont {V.~W.}\ \bibnamefont {Brar}},
  \bibinfo {author} {\bibfnamefont {A.~V.}\ \bibnamefont {Shytov}}, \bibinfo
  {author} {\bibfnamefont {Q.}~\bibnamefont {Wu}}, \bibinfo {author}
  {\bibfnamefont {W.}~\bibnamefont {R.}}, \bibinfo {author} {\bibfnamefont
  {H.-Z.}\ \bibnamefont {Tsai}}, \bibinfo {author} {\bibfnamefont
  {A.}~\bibnamefont {Zettl}}, \bibinfo {author} {\bibfnamefont {L.~S.}\
  \bibnamefont {Levitov}}, \ and\ \bibinfo {author} {\bibfnamefont {M.~F.}\
  \bibnamefont {Crommie}},\ }\href {\doibase 10.1038/nphys2379} {\bibfield
  {journal} {\bibinfo  {journal} {Nature Physics}\ }\textbf {\bibinfo {volume}
  {8}},\ \bibinfo {pages} {653} (\bibinfo {year} {2012})}\BibitemShut {NoStop}%
\bibitem [{\citenamefont {Wang}\ \emph {et~al.}(2013)\citenamefont {Wang},
  \citenamefont {Wong}, \citenamefont {Shytov}, \citenamefont {Brar},
  \citenamefont {Choi}, \citenamefont {Wu}, \citenamefont {Tsai}, \citenamefont
  {Regan}, \citenamefont {Zettl}, \citenamefont {Kawakami}, \citenamefont
  {Louie}, \citenamefont {Levitov},\ and\ \citenamefont {Crommie}}]{Wang2013}%
  \BibitemOpen
  \bibfield  {author} {\bibinfo {author} {\bibfnamefont {Y.}~\bibnamefont
  {Wang}}, \bibinfo {author} {\bibfnamefont {D.}~\bibnamefont {Wong}}, \bibinfo
  {author} {\bibfnamefont {A.~V.}\ \bibnamefont {Shytov}}, \bibinfo {author}
  {\bibfnamefont {V.~W.}\ \bibnamefont {Brar}}, \bibinfo {author}
  {\bibfnamefont {S.}~\bibnamefont {Choi}}, \bibinfo {author} {\bibfnamefont
  {Q.}~\bibnamefont {Wu}}, \bibinfo {author} {\bibfnamefont {H.-Z.}\
  \bibnamefont {Tsai}}, \bibinfo {author} {\bibfnamefont {W.}~\bibnamefont
  {Regan}}, \bibinfo {author} {\bibfnamefont {A.}~\bibnamefont {Zettl}},
  \bibinfo {author} {\bibfnamefont {R.~K.}\ \bibnamefont {Kawakami}}, \bibinfo
  {author} {\bibfnamefont {S.~G.}\ \bibnamefont {Louie}}, \bibinfo {author}
  {\bibfnamefont {L.~S.}\ \bibnamefont {Levitov}}, \ and\ \bibinfo {author}
  {\bibfnamefont {M.~F.}\ \bibnamefont {Crommie}},\ }\href {\doibase
  10.1126/science.1234320} {\bibfield  {journal} {\bibinfo  {journal}
  {Science}\ }\textbf {\bibinfo {volume} {340}},\ \bibinfo {pages} {734}
  (\bibinfo {year} {2013})}\BibitemShut {NoStop}%
\bibitem [{\citenamefont {Wang}\ \emph {et~al.}(2015)\citenamefont {Wang},
  \citenamefont {Xiao}, \citenamefont {Cai}, \citenamefont {Bao}, \citenamefont
  {Reutt-Robey},\ and\ \citenamefont {Fuhrer}}]{Wang2015}%
  \BibitemOpen
  \bibfield  {author} {\bibinfo {author} {\bibfnamefont {Y.}~\bibnamefont
  {Wang}}, \bibinfo {author} {\bibfnamefont {S.}~\bibnamefont {Xiao}}, \bibinfo
  {author} {\bibfnamefont {X.}~\bibnamefont {Cai}}, \bibinfo {author}
  {\bibfnamefont {W.}~\bibnamefont {Bao}}, \bibinfo {author} {\bibfnamefont
  {J.}~\bibnamefont {Reutt-Robey}}, \ and\ \bibinfo {author} {\bibfnamefont
  {M.~S.}\ \bibnamefont {Fuhrer}},\ }\href {\doibase 10.1038/srep15764}
  {\bibfield  {journal} {\bibinfo  {journal} {Scientific Reports}\ }\textbf
  {\bibinfo {volume} {5}} (\bibinfo {year} {2015}),\
  10.1038/srep15764}\BibitemShut {NoStop}%
\bibitem [{\citenamefont {Mao}\ \emph {et~al.}(2016)\citenamefont {Mao},
  \citenamefont {Jiang}, \citenamefont {Moldovan}, \citenamefont {Li},
  \citenamefont {Watanabe}, \citenamefont {Taniguchi}, \citenamefont {Masir},
  \citenamefont {Peeters},\ and\ \citenamefont {Andrei}}]{Mao2016}%
  \BibitemOpen
  \bibfield  {author} {\bibinfo {author} {\bibfnamefont {J.}~\bibnamefont
  {Mao}}, \bibinfo {author} {\bibfnamefont {Y.}~\bibnamefont {Jiang}}, \bibinfo
  {author} {\bibfnamefont {D.}~\bibnamefont {Moldovan}}, \bibinfo {author}
  {\bibfnamefont {G.}~\bibnamefont {Li}}, \bibinfo {author} {\bibfnamefont
  {K.}~\bibnamefont {Watanabe}}, \bibinfo {author} {\bibfnamefont
  {T.}~\bibnamefont {Taniguchi}}, \bibinfo {author} {\bibfnamefont {M.~R.}\
  \bibnamefont {Masir}}, \bibinfo {author} {\bibfnamefont {F.~M.}\ \bibnamefont
  {Peeters}}, \ and\ \bibinfo {author} {\bibfnamefont {E.~Y.}\ \bibnamefont
  {Andrei}},\ }\href {\doibase 10.1038/nphys3665} {\bibfield  {journal}
  {\bibinfo  {journal} {Nature Physics}\ }\textbf {\bibinfo {volume} {12}},\
  \bibinfo {pages} {545} (\bibinfo {year} {2016})}\BibitemShut {NoStop}%
\bibitem [{\citenamefont {Brar}\ \emph {et~al.}(2011)\citenamefont {Brar},
  \citenamefont {Decker}, \citenamefont {Solowan}, \citenamefont {Wang},
  \citenamefont {Maserati}, \citenamefont {Chan}, \citenamefont {Lee},
  \citenamefont {Girit}, \citenamefont {Zettl}, \citenamefont {Louie},
  \citenamefont {Cohen},\ and\ \citenamefont {Crommie}}]{Brar2011}%
  \BibitemOpen
  \bibfield  {author} {\bibinfo {author} {\bibfnamefont {V.~W.}\ \bibnamefont
  {Brar}}, \bibinfo {author} {\bibfnamefont {R.}~\bibnamefont {Decker}},
  \bibinfo {author} {\bibfnamefont {H.-M.}\ \bibnamefont {Solowan}}, \bibinfo
  {author} {\bibfnamefont {Y.}~\bibnamefont {Wang}}, \bibinfo {author}
  {\bibfnamefont {L.}~\bibnamefont {Maserati}}, \bibinfo {author}
  {\bibfnamefont {K.~T.}\ \bibnamefont {Chan}}, \bibinfo {author}
  {\bibfnamefont {H.}~\bibnamefont {Lee}}, \bibinfo {author} {\bibfnamefont
  {.~O.}\ \bibnamefont {Girit}}, \bibinfo {author} {\bibfnamefont
  {A.}~\bibnamefont {Zettl}}, \bibinfo {author} {\bibfnamefont {S.~G.}\
  \bibnamefont {Louie}}, \bibinfo {author} {\bibfnamefont {M.~L.}\ \bibnamefont
  {Cohen}}, \ and\ \bibinfo {author} {\bibfnamefont {M.~F.}\ \bibnamefont
  {Crommie}},\ }\href {\doibase 10.1038/nphys1807} {\bibfield  {journal}
  {\bibinfo  {journal} {Nature Physics}\ }\textbf {\bibinfo {volume} {7}},\
  \bibinfo {pages} {43} (\bibinfo {year} {2011})}\BibitemShut {NoStop}%
\bibitem [{\citenamefont {Decker}\ \emph {et~al.}(2011)\citenamefont {Decker},
  \citenamefont {Wang}, \citenamefont {Brar}, \citenamefont {Regan},
  \citenamefont {Tsai}, \citenamefont {Wu}, \citenamefont {Gannett},
  \citenamefont {Zettl},\ and\ \citenamefont {Crommie}}]{Decker2011}%
  \BibitemOpen
  \bibfield  {author} {\bibinfo {author} {\bibfnamefont {R.}~\bibnamefont
  {Decker}}, \bibinfo {author} {\bibfnamefont {Y.}~\bibnamefont {Wang}},
  \bibinfo {author} {\bibfnamefont {V.~W.}\ \bibnamefont {Brar}}, \bibinfo
  {author} {\bibfnamefont {W.}~\bibnamefont {Regan}}, \bibinfo {author}
  {\bibfnamefont {H.-Z.}\ \bibnamefont {Tsai}}, \bibinfo {author}
  {\bibfnamefont {Q.}~\bibnamefont {Wu}}, \bibinfo {author} {\bibfnamefont
  {W.}~\bibnamefont {Gannett}}, \bibinfo {author} {\bibfnamefont
  {A.}~\bibnamefont {Zettl}}, \ and\ \bibinfo {author} {\bibfnamefont {M.~F.}\
  \bibnamefont {Crommie}},\ }\href {\doibase 10.1021/nl2005115} {\bibfield
  {journal} {\bibinfo  {journal} {Nano Letters}\ }\textbf {\bibinfo {volume}
  {11}},\ \bibinfo {pages} {2291} (\bibinfo {year} {2011})},\ \bibinfo {note}
  {pMID: 21553853}\BibitemShut {NoStop}%
\bibitem [{\citenamefont {Burson}\ \emph {et~al.}(2013)\citenamefont {Burson},
  \citenamefont {Cullen}, \citenamefont {Adam}, \citenamefont {Dean},
  \citenamefont {Watanabe}, \citenamefont {Taniguchi}, \citenamefont {Kim},\
  and\ \citenamefont {Fuhrer}}]{Burson2013}%
  \BibitemOpen
  \bibfield  {author} {\bibinfo {author} {\bibfnamefont {K.~M.}\ \bibnamefont
  {Burson}}, \bibinfo {author} {\bibfnamefont {W.~G.}\ \bibnamefont {Cullen}},
  \bibinfo {author} {\bibfnamefont {S.}~\bibnamefont {Adam}}, \bibinfo {author}
  {\bibfnamefont {C.~R.}\ \bibnamefont {Dean}}, \bibinfo {author}
  {\bibfnamefont {K.}~\bibnamefont {Watanabe}}, \bibinfo {author}
  {\bibfnamefont {T.}~\bibnamefont {Taniguchi}}, \bibinfo {author}
  {\bibfnamefont {P.}~\bibnamefont {Kim}}, \ and\ \bibinfo {author}
  {\bibfnamefont {M.~S.}\ \bibnamefont {Fuhrer}},\ }\href {\doibase
  10.1021/nl4012529} {\bibfield  {journal} {\bibinfo  {journal} {Nano Letters}\
  }\textbf {\bibinfo {volume} {13}},\ \bibinfo {pages} {3576} (\bibinfo {year}
  {2013})},\ \bibinfo {note} {pMID: 23879288}\BibitemShut {NoStop}%
\bibitem [{\citenamefont {Dean}\ \emph {et~al.}(2010)\citenamefont {Dean},
  \citenamefont {Young}, \citenamefont {Meric}, \citenamefont {Lee},
  \citenamefont {Wang}, \citenamefont {Sorgenfrei}, \citenamefont {Watanabe},
  \citenamefont {Taniguchi}, \citenamefont {Kim}, \citenamefont {Shepard},\
  and\ \citenamefont {Hone}}]{Dean2010}%
  \BibitemOpen
  \bibfield  {author} {\bibinfo {author} {\bibfnamefont {C.~R.}\ \bibnamefont
  {Dean}}, \bibinfo {author} {\bibfnamefont {A.~F.}\ \bibnamefont {Young}},
  \bibinfo {author} {\bibfnamefont {I.}~\bibnamefont {Meric}}, \bibinfo
  {author} {\bibfnamefont {C.}~\bibnamefont {Lee}}, \bibinfo {author}
  {\bibfnamefont {L.}~\bibnamefont {Wang}}, \bibinfo {author} {\bibfnamefont
  {S.}~\bibnamefont {Sorgenfrei}}, \bibinfo {author} {\bibfnamefont
  {K.}~\bibnamefont {Watanabe}}, \bibinfo {author} {\bibfnamefont
  {T.}~\bibnamefont {Taniguchi}}, \bibinfo {author} {\bibfnamefont
  {P.}~\bibnamefont {Kim}}, \bibinfo {author} {\bibfnamefont {K.~L.}\
  \bibnamefont {Shepard}}, \ and\ \bibinfo {author} {\bibfnamefont
  {J.}~\bibnamefont {Hone}},\ }\href {\doibase 10.1038/nnano.2010.172}
  {\bibfield  {journal} {\bibinfo  {journal} {Nature Nanotechnology}\ }\textbf
  {\bibinfo {volume} {5}},\ \bibinfo {pages} {722} (\bibinfo {year}
  {2010})}\BibitemShut {NoStop}%
\bibitem [{\citenamefont {Rengel}\ \emph {et~al.}(2015)\citenamefont {Rengel},
  \citenamefont {Iglesias}, \citenamefont {Pascual},\ and\ \citenamefont
  {Martin}}]{Rengel2015}%
  \BibitemOpen
  \bibfield  {author} {\bibinfo {author} {\bibfnamefont {R.}~\bibnamefont
  {Rengel}}, \bibinfo {author} {\bibfnamefont {J.~M.}\ \bibnamefont
  {Iglesias}}, \bibinfo {author} {\bibfnamefont {E.}~\bibnamefont {Pascual}}, \
  and\ \bibinfo {author} {\bibfnamefont {M.~J.}\ \bibnamefont {Martin}},\
  }\href {\doibase 10.1088/1742-6596/647/1/012046} {\bibfield  {journal}
  {\bibinfo  {journal} {Journal of Physics: Conference Series}\ }\textbf
  {\bibinfo {volume} {647}},\ \bibinfo {pages} {012046} (\bibinfo {year}
  {2015})}\BibitemShut {NoStop}%
\end{thebibliography}%

\end{document}